\documentclass[pdflatex,sn-mathphys-num]{sn-jnl}
\usepackage{graphicx}%
\usepackage{multirow}%
\usepackage{amsmath,amssymb,amsfonts}%
\usepackage{amsthm}%
\usepackage{mathrsfs}%
\usepackage[title]{appendix}%
\usepackage{xcolor}%
\definecolor{ForestGreen}{rgb}{0.13, 0.55, 0.13}
\usepackage{textcomp}%
\usepackage{manyfoot}%
\usepackage{booktabs}%
\usepackage{algorithm}%
\usepackage{algorithmicx}%
\usepackage{algpseudocode}%
\usepackage{listings}%
\usepackage{url}
\usepackage{multirow}
\usepackage{enumitem}
\usepackage{array}  
\usepackage{booktabs} 

\usepackage{longtable}
\usepackage{array}
\usepackage{ragged2e}
\usepackage{geometry}
\usepackage{caption}
\usepackage{adjustbox}
\usepackage{booktabs} % for better table lines

\usepackage{siunitx}
\usepackage[most]{tcolorbox}

\newtcolorbox{summarybox}{
  enhanced,
  colback=gray!5,
  colframe=gray!50!black,
  rounded corners,
}

\begin{document}

\title[Article Title]{A Multimodal Approach Combining Biometrics and Self-Report Instruments for Monitoring Stress in Programming: Methodological Insights}

% going to https://emsejournal.github.io/special_issues/2024_SI_RENE.html 
\author*[1]{\fnm{Cristina} \sur{Martinez Montes}
{\href{https://orcid.org/0000-0003-1150-6931}{\includegraphics[scale=0.9]{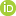}}}
}\email{montesc@chalmers.se}

\author[2]{\fnm{Daniela} \sur{Grassi} 
{\href{https://orcid.org/0000-0001-8376-1232}{\includegraphics[scale=0.9]{ORCIDiD_icon16x16.png}}}
}
\email{daniela.grassi@uniba.it}
% https://orcid.org/0000-0001-8376-1232
% Birgit: Couldn't find ORCID for Daniela, but we can add later

\author[2]{\fnm{Nicole} \sur{Novielli}
{\href{https://orcid.org/0000-0003-1160-2608}{\includegraphics[scale=0.9]{ORCIDiD_icon16x16.png}}}
}\email{nicole.novielli@uniba.it}

\author[1,3]{\fnm{Birgit} \sur{Penzenstadler}{\href{https://orcid.org/0000-0002-5771-0455}{\includegraphics[scale=0.9]{ORCIDiD_icon16x16.png}}}}\email{birgitp@chalmers.se}

\affil*[1]{\orgdiv{Computer Science and Engineering}, \orgname{Chalmers University of Technology and University of Gothenburg}, \orgaddress \city{Gothenburg}, \postcode{41756}, \country{Sweden}}

\affil[2]{\orgdiv{Computer Science}, \orgname{University of Bari}, \postcode{70125}, \country{Italy}}

\affil[3]{\orgname{Lappeenranta University of Technology}, \orgaddress{\city{Lappeenranta}, \country{Finland}}}

%\renewcommand{\shortauthors}{Anonymous}

\abstract{
  
The study of well-being, stress and other human factors has traditionally relied on self-report instruments to assess key variables. However, concerns about potential biases in these instruments, even when thoroughly validated and standardised, have driven growing interest in alternatives in combining these measures with more objective methods such as physiological measures.

We aimed to (i) compare psychometric stress measures and biometric indicators and (ii) identify stress-related patterns in biometric data during software engineering tasks.

 We conducted an experiment where participants completed a pre-survey, then programmed two tasks wearing biometric sensors, answered brief post-surveys for each, and finally went through a short exit interview.

Our results showed diverse outcomes; we found no stress in the psychometric instruments. Participants in the interviews reported a mix of feeling no stress and experiencing time pressure. Finally, the biometrics showed a significant difference only in EDA phasic peaks.

We conclude that our chosen way of inducing stress by imposing a stricter time limit was insufficient. We offer methodological insights for future studies working with stress, biometrics, and psychometric instruments.
}



\maketitle

\section{Introduction} \label{sec:intro}

%background

Software engineering (SE) is a cognitively demanding profession that requires intense focus, problem-solving, and creativity. However, these tasks often come with high-stress levels due to the work characteristics, which often involve long working hours, high cognitive load, frequent interruptions, task interdependence and tight deadlines~\cite{godliauskas2025well}. Prolonged exposure to such stressors can lead to burnout, a state of emotional, mental, and physical exhaustion that negatively impacts individual well-being and organisational productivity \cite{maslach2001job}. 

Understanding the stressors specific to SE tasks and accurately measuring their impact is essential for developing effective interventions to mitigate these risks.
%Problem Statement
Research on emotions, affect, and stress in software engineering has mainly used self-reported instruments, such as surveys, interviews and psychometric instruments. Graziotin et al., 2014 \cite{graziotin2014happy}, were among the first researchers in the area proposing to study human factors using psychological measurements. Studies on happiness \cite{graziotin2019happiness}, attention awareness \cite{bernardez2023empirical}, positive and negative experience, psychological well-being \cite{montes2024qualifying, montes2023piloting}, positive thinking, and self-efficacy \cite{penzenstadler2022take} have been conducted using psychometric instruments to assess these constructs. 

While these methods offer insights into subjective experiences, they are prone to biases, including recall bias, social desirability bias (SDR), and acquiescent responding (ACQ) \cite{kreitchmann2019controlling}. SDR refers to the tendency to respond in a way consistent with what is perceived as desirable by salient others \cite{kuncel2009conceptual}. Meanwhile, ACQ relates to the tendency to favour the positive end of the rating scale, irrespective of the item's content \cite{weijters2013reversed}.

Additionally, self-reported measures may not fully capture stress's physiological and cognitive aspects, essential for understanding its impact on performance and well-being. To address these limitations, recent studies have investigated the use of biometrics to recognise developers' emotions during programming tasks~\cite{MF15,Vrzakova2020,girardi2020recognizing,Grassi_etAl:2025}. What these studies have in common is the operationalisation of emotions along the dimension of valence, i.e. the (un)pleasantness of the emotional stimulus, and arousal, i.e. the level of emotional activation~\cite{russell1980circumplex}, showing promising results in their recognition through machine-learning supervised classifiers. 

Despite advances in this domain, the literature reveals a significant gap, as, to the best of our knowledge, no research has specifically addressed stress. As a result, there is a growing need for more objective and reliable methods to assess stress in software engineering contexts. At the same time, recent work by Westerink and colleagues~\cite{westerink2020deriving} provided empirical evidence that biometrics collected with non-invasive sensors can be used as a stress indicator.
%, with changes in electrodermal activity (EDA) positively correlating with changes in cortisol, which is the hormone associated with the experience of stress. 
Inspired by these findings, we decided to perform an empirical study to fill this gap, towards enhancing the accuracy and reliability of stress measurement in software engineering. This decision was in line with our long-term goal to support early detection of stress, thus enabling interventions to prevent its long-term negative effects on well-being and productivity.

We designed and implemented an empirical study with the primary goal of investigating to what extent we can use biometrics as a proxy for stress experienced by software developers during programming tasks, to reduce the reliance on self-reported data and obtain a more comprehensive understanding of stress related to SE tasks. To this aim, we compare biometric measurements with traditional psychometric instruments as collected during programming tasks performed by ten developers in a controlled lab environment. 
Although we invested considerable time and resources in the design of the empirical protocol, we obtained disappointing outcomes due to the inability to induce stress in the participants of our empirical study. This prompted us to redirect our efforts toward a comprehensive assessment of the robustness of the protocol we adopted, thus deriving methodological guidelines to inform future studies on this topic. 

A key finding was that the intended stress manipulation through time pressure failed to produce measurable stress responses at the group level. This led us to conclude that time pressure alone may be insufficient to induce stress in experienced programmers. Future studies should consider multi-stressor approaches or tasks with higher personal stakes for participants.
Furthermore, our findings reveal that the individual-level triangulation of data sources provided more nuanced insights than the group level. This can be observed by the combined analysis of self-reported stress measures,  electrodermal activity (EDA) peaks, and qualitative interview data on a participant-by-participant basis. 
%In particular, this multi-modal analysis shows moderate alignment between these different measurement approaches for individual participants, though not perfect correlations. For instance, some participants showed consistent stress levels across all conditions, which aligned with their interview responses describing mild frustration rather than significant stress. Others exhibited pre-task stress that decreased during actual coding activities, suggesting that anticipation of participating in the study was more stressful than the programming tasks themselves.
Finally, we discuss methodological challenges associated with distinguishing between acute and chronic stress, which might be a confounder in a lab setting focusing on stress detection during coding tasks. Specifically, we noted that while our multi-modal measurement approach showed sensitivity to stress variations, the ethical constraints of inducing stress in research settings may fundamentally limit the ability to create strong enough stressors that eventually yield actionable data without crossing ethical boundaries.

The remainder of the paper is organised as follows. In Section \ref{sec:backgr} , we present the background and discuss the related work on stress and biometrics in software engineering. Then, in Section \ref{sec:methodology} we describe the methodology, including the experimental protocol for data collection and the method of analysis of psychometrics, biometrics, and interviews with participants. Results are presented in Section \ref{sec:results} and discussed in Section \ref{sec:discussion}, where we also present the threats to validity and the strategies adopted to mitigate them. Finally, we conclude the paper and discuss future work directions in Section \ref{sec:conclusion}.
%Goal

%\begin{itemize}
%    \item \textbf {Primary Goal:} to compare biometric measurements (such as electroencephalography (EEG) and electrodermal activity (EDA)) with traditional psychometric instruments to reduce the reliance on self-reported data and obtain a more comprehensive understanding of stress related to SE tasks.
%\end{itemize}

%\textbf{Secondary Goals:}
%\begin{enumerate}[label=\alph*)]
%    \item Determine whether stress measurements from one device (e.g., EEG) can replace another (e.g., EDA or psychometrics) or whether a particular device is less reliable in specific contexts.
%    \item Measure stress levels in real-time during SE tasks to identify patterns and triggers of stress in a naturalistic work environment.
%\end{enumerate}

%importance

\section{Background and Related Work} \label{sec:backgr}

%To add a glossary

Physiological measures, such as electroencephalography (EEG), electrodermal activity (EDA), and heart-related metrics, have become valuable tools for studying cognitive load and stress across various domains. These measures offer objective insights into mental states, offering advantages over traditional self-reported methods. In fact, biometrics hold the potential to address the limitations of self-report methods by providing objective, continuous measurement of the biometric changes that are induced by mental states \cite{Kocielnik}. 
Among other affective states, in this study we specifically focus on the study of stress, that is, the physiological or psychological response to internal or external triggers, involving people's bodily reactions, feelings and behaviour (see Table \ref{tab:definitions}). 
In the bi-dimensional categorisation of emotions along the concepts of valence and arousal, stress is positioned in the scope of negative emotions~\cite{fan:stress:anxiety} and associated with high arousal \cite{Russell2005}. This positioning reflects the nature of stress as an unpleasant emotional state that involves high physiological and psychological activation. Stress appears near other similar emotional states such as anxiety, tension, distress, and nervousness in this model.
In the following, we report foundational related work on the use of biometrics for the study of cognitive and emotional states (Section \ref{sec:background}). We complement this background knowledge with an overview of recent related studies in the field of software engineering  (Section \ref{sec:relatedwork}).

Table \ref{tab:definitions} presents definitions for the most important concepts in this study.

\begin{table}[h]
\centering
\caption{Operationalisation of main concepts based on the American Psychological Association definitions \cite{apa2018dictionary}}
\label{tab:definitions}
\begin{tabular}{l p{10cm}} 
\toprule
\textbf{Concept} & \textbf{Definition} \\
\midrule
Stress & ``The physiological or psychological response to internal or external stressors. Stress involves changes affecting nearly every system of the body, influencing how people feel and behave." \\
%Emotion & ``A complex reaction pattern, involving experiential, behavioural, and physiological elements, by which an individual attempts to deal with a personally significant matter or event." \\
Mental Workload & ``The relative demand imposed by a particular task, in terms of mental resources required." \\
\bottomrule
\end{tabular}
\end{table}

\subsection{\textbf{Physiological Measures of Stress and Mental Load}}
\label{sec:background}

\textbf{Stress}. The link between affective states and physiological feedback, collected with biometric sensors, has been investigated for a long time by researchers in the affective computing community \cite{KimA08, kim2004emotion, soleymani2015analysis, KoelstraMSLYEPNP12}. In recent years, the study of emotions and their recognition has gained attentions also in software engineering research, due to their influence on developers’ wellbeing, stress levels, and cognitive performance  \cite{graziotin2014happy,Khan_debug_emotion}.

%\nicole{This paragraph about emotion modelling doesn't fit well here} To represent emotions, our study adopts the Circumplex Model of Affect \cite{. https://doi.org/10.1037/h0077714}, a dimensional framework that positions emotions within a two-dimensional space defined by valence (ranging from unpleasant to pleasant) and arousal (ranging from calm to excited) \cite{. https://doi.org/10.1037/h0077714}. This model is particularly suited to software engineering contexts, as it avoids cultural and linguistic biases commonly associated with discrete emotion models \cite{https://doi.org/10.1037/0033-2909.110.3.426}. 

Various biometric signals have been employed to detect affective states. In particular, EEG has been widely used to analyse changes in brain activity correlated with emotional valence (pleasant vs. unpleasant emotional stimulus) and arousal (i.e., high vs. low level of emotional activation) \cite{Russell1980}. For instance, high-frequency bands such as gamma have shown strong correlations with valence, particularly in the frontal and parietal lobes \cite{soleymani2015analysis}. EEG also enables computation of Frontal Alpha Asymmetry, a known biomarker linked to emotional valence and stress \cite{goodman2013stress} %\daniela{check this references}\nicole{@Daniela: do not forget to add the missing references here}.
Moreover, EDA is widely adopted due to its association with the arousal dimension \cite{bradley2000measuring}. EDA has thus been effectively used to identify emotions \cite{TSE_Girardi, girardi2020recognizing}. Its sensitivity to emotional intensity makes it a valuable, non-invasive proxy for monitoring real-time emotional fluctuations during cognitive tasks.
Furthermore, HR and HRV metrics also provide insights into emotional arousal and cognitive load. Specifically, HRV indicators such as RMSSD and LF/HF ratio have been shown to reflect sympathetic and parasympathetic nervous system activity, which are modulated during emotional and stress responses \cite{Canento_2011, castaldo2015acute}

Similarly to what was done for the recognition of emotions, the study of biometrics has been applied to the recognition of stress episodes. In particular, EEG has been used to identify specific brainwave patterns, such as alpha and beta frequencies, which are closely linked to stress. In their study, Saeed et al. \cite{saeed2020eeg} found that alpha asymmetry could be a potential reliable biomarker for stress classification. They complemented the EEG data with the Perceived Stress Scale (PSS-10)  and an interview to obtain a thorough understanding of stress. A similar setup was used in our study to get a more complete view of how stress manifests physiologically and emotionally.
A similar study by Chae et al. \cite{chae2021relationship} looked at the relationship between stress levels and rework using EEG, EDA, and a survey, finding that all three measures consistently indicated that rework caused stress in workers. They emphasised that excessive occupational stress can negatively affect employee work performance and work-life balance. Additionally, they stressed the cognitive and emotional toll of repetitive tasks. Our study builds on this by comparing stress measurements from EEG, EDA, and psychometric instruments to enhance the understanding of workplace stress, particularly in high-pressure environments such as software engineering.

As for EDA, its link with stress episodes was demonstrated by Westerink et al.~\cite{westerink2020deriving}. They explored the use of physiological sensors for detecting stress episodes. Their findings revealed a significant relationship between cortisol fluctuations (the primary stress hormone) and electrodermal activity (EDA) measurements. Notably, peaks in skin conductance preceded cortisol elevations, which suggests that EDA monitoring could serve as an early warning system for stress onset.
In related work, Kocielnik et al.~\cite{Kocielnik} developed an approach that integrated EDA measurements with calendar data to examine potential connections between daily activities and stress responses. The paper presents a framework for long-term, unobtrusive stress monitoring in workplace settings using a wearable sensor wristband (DTI-2) that measures skin conductivity. The authors proposed an approach to process EDA raw signals to identify stress levels and visualise this data in relation to users' calendar activities. Through field studies with university staff, they demonstrated that this approach helps users discover meaningful stress patterns they weren't previously aware of.

\textbf{Mental Workload}. Linked to stress and based on the premise that workload affects performance, Mohanavelu et al.'s \cite{mohanavelu2020dynamic} study focused on measuring and understanding the cognitive workload and attention during different levels of task difficulty: normal, moderate, high, and very high workloads. They used EEG to track how the brain responds under varying workloads and a NASA-Task Load Index (NASA-TLX) questionnaire to validate their findings. Results from EEG showed that the prefrontal, frontotemporal, and parietal brain regions were highly engaged under high and very high workloads; NASA-TLX results aligned well with EEG data.
Considering the previous results and since software developers' work also demands a high mental workload, we used NASA TLX in this study to capture subjective workload data.

Similar to the previous study and considering sleep deprivation, which is also quite present in the software engineering field, Martínez Vásquez et al. \cite{martinez2023mutual} collected data from ten participants performing cognitive tasks every two hours for 24 hours to explore the relationship between brain activity (EEG) and autonomic sympathetic activity (EDA) under sleep deprivation, aiming to assess their role in determining readiness for cognitive tasks. Based on their findings, the authors proposed that the mutual information between EDA and EEG signals reported in their study indicates that examining EDA could offer a compelling alternative for studying brain activity. In our study, we collected both data to explore their relationship with cognitive and emotional processes.

\subsection{Using Biometrics for Studying Cognitive and Affective States in Software Development}
%\daniela{Maybe Biometrics Studies in Software Development Tasks? Since we are not focusing only on EEG}
\label{sec:relatedwork}

Researchers in software engineering have explored connections between developers' cognitive states---as measured through physiological indicators---and various software development dimensions, including comprehension of code~\cite{Fucci_etAL:ICPC2019,fucci2019replication}, developer productivity and interruptibility~\cite{RHM15,ZCM17}, and the emotions experienced by developers during programming tasks~\cite{MF15,girardi2020recognizing,Vrzakova2020,TSE_Girardi}.

%Recent research has explored the application of physiological measurements to study software developers' productivity. 
\textbf{EEG}. Several studies have been done to research brain activity during programming tasks, focusing on the cognitive load and mental effort involved in software development. For instance, Calcagno et al. \cite {calcagno2020eeg} investigated brain activity during programming tasks using EEG with ten experienced software developers. Their results showed significant changes in brain activity when transitioning from a baseline condition (typing with eyes closed) to a programming task. Specifically, they observed a decrease in Alpha power and an increase in Delta, Theta, and Beta power, particularly in the frontal and parieto-occipital regions. The increase in Beta activity was most prominent at the beginning of the task, likely reflecting the heightened alertness and attention required for understanding instructions and planning code implementation. In contrast, Theta and Delta power increased during later phases, suggesting greater mental workload and working memory engagement. Their results suggest that EEG measures can provide insights into cognitive load and attentional dynamics during software development tasks.

Medeiros et al. \cite {medeiros2021can} performed a controlled experiment on task comprehension with 26 programmers using three code snippets in Java with different complexity levels. The study found that features related to Theta, Alpha, and Beta brain waves were the most effective at identifying levels of mental effort required by different code lines. The EEG data indicated signs of mental effort saturation as code complexity increased. In contrast, traditional software complexity metrics did not accurately reflect the cognitive effort required for code comprehension. 

Radevski et al.~\cite{RHM15} introduced a framework that continuously monitors developers' productivity by tracking electrical activity in the brain, to assess and improve their productivity. Their proposed approach relies on off-the-shelf EEG devices to support their long-term goal of detecting negative cognitive and emotional states such as stress, fatigue, and frustration, which might emerge during programming tasks. While not being assessed for the specific task of emotion detection, the framework's usability was evaluated through a pilot user study with six participants who wore the device for an entire workday, finding it was feasible but had some comfort issues. Their study also addresses ethical considerations and user acceptance challenges that must be considered when conducting empirical studies involving the use of biometric devices. 

\textbf{Combining data from multiple sensors}. Beyond EEG, various approaches have been proposed based on different sensor combinations for the recognition of emotional and cognitive states. 
M\"uller and Fritz~\cite{MF15} employed a combination of biometric indicators to assess both progress and interruptibility during small development tasks. They demonstrated that emotional states of developers during programming tasks could be classified with 71\% accuracy by analysing a rich set of physiological signals, including brainwave frequencies, pupil dimensions, and heart rate. They also report achieving a comparable accuracy when predicting developers' self-perceived progress during development tasks, though this required a distinct set of biometric indicators encompassing EDA signals, skin temperature, brainwave patterns, and pupil size variations.

In a partial replication of the original study by M\"uller and Fritz~\cite{MF15}, Girardi et al. conducted an empirical investigation to identify the minimal configuration of non-invasive biometric sensors for recognizing emotions during programming tasks~\cite{girardi2020recognizing}. They developed two supervised classification models for valence and arousal dimensions using emotions self-reported by 23 participants during a Java programming assignment as a ground truth.
Through experimentation with various biometric combinations, they found that developers' emotional valence and arousal could be reliably detected using a combination of electrodermal activity (EDA) and heart-related measurements, collected via the Empatica E4 wristband, suitable for emotion detection during software development activities. 
Using only the wristband, they achieved accuracy levels for valence ($.71$) and arousal ($.65$) comparable to those obtained with the complete sensor array (wristband + EEG helmet). Consequently, in their subsequent study, they utilized only the Empatica wristband for measuring both electrodermal activity and heart-related biometric signals~\cite{TSE_Girardi}.
Their study not only confirmed previous findings by M"uller and Fritz~\cite{MF15} regarding non-invasive sensors' reliability for valence classification, but also extended this work by developing an arousal dimension classifier.
%The researchers identified a minimal sensor configuration—EDA, BVP, and HR measurements collected via the Empatica E4 wristband—suitable for emotion detection during software development activities. 

Vrzakova et al.~\cite{Vrzakova2020} combined eye tracking measurements and electrodermal activity to classify emotional valence and arousal of software developers during code review activities. They conducted an \textit{in-situ} study with 37 professional developers engaged in code review tasks.
They used features extracted from individual signal types as well as combined feature sets incorporating all available signals to train supervised machine learning models. For evaluation, they established a ground truth using binarised self-reported emotional scores for valence (positive vs.) negative and arousal (low vs. high).
Their findings revealed that eye gaze measurements provided the strongest predictive capability for both emotional dimensions, achieving accuracy rates of 85.8\% for valence and 76.6\% for arousal. However, when incorporating features from all physiological signals, including EDA, in their supervised models, they observed a boost of classification performance for both valence and arousal dimensions, with accuracy rates reaching 90.7\% and 83.9\%, respectively.

\section{Methodology} \label{sec:methodology}

We designed an experiment with two main objectives: (1) to identify the stress levels induced by programming tasks and (2) to evaluate the accuracy of self-reported instruments in measuring stress in comparison with biometric measurements. To achieve this, we investigated the following questions:

\begin{enumerate}
    \item How reliable are psychometric stress measures compared to real-time biometric indicators (EEG and EDA) during software engineering tasks? 
    \item What stress-related patterns can be identified in real-time biometric data (EEG and EDA) during software engineering tasks?
\end{enumerate}

This section outlines the experimental protocol, the instruments used for data collection, and the approach to data analysis.

\begin{figure*}
    \centering
    \includegraphics[width=1\linewidth]{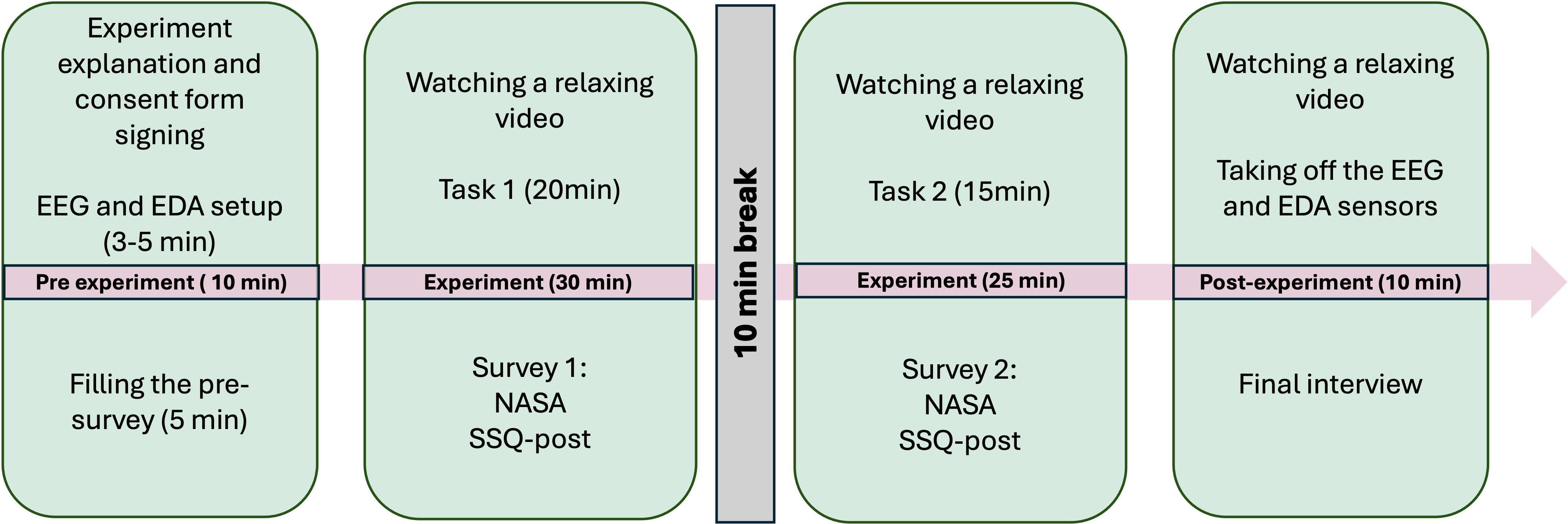}
    \caption{Experiment's Timeline}
    \label{fig:timeline}
\end{figure*}

\subsection{Participants and Recruitment Strategy}

The study included ten participants: nine PhD students in computer science, artificial intelligence, and bioinformatics and one master's student in data science. They were in various academic stages, from the first to third year, and reported confidence levels in programming ranging from ``somewhat confident" to ``very confident." Python was the preferred programming language for most, except one who preferred Java.

\subsection{Experiment Setup}

To conduct the experiment, we recruited subjects from the Ph.D. and master's students in Computer Science who could code in Python or Java. We collected the preliminary availability of volunteers and scheduled the experimental sessions based on their agendas over a time span of two weeks. 

\textbf{Pre-experimental Briefing.}
Participants began by listening to the explanation of the experiment, reading the informed consent form, and having the opportunity to ask questions. After signing the informed consent form, participants wore biometric sensors, and the researchers made sure the signals were being captured correctly and started the recording. Subsequently, participants completed the first survey (PSS-10, and SSSQ-pre) and then watched a two-minute relaxation video to induce relaxation and establish a neutral emotional state \cite{coan2007handbook}. The signal collected in this neutral emotional state is used as a baseline for each participant, which is required to preprocess the raw biometric signal, as explained in Section \ref{sec:biometric_analysis}.

\textbf{Programming Tasks and Data Collection.} Participants started carrying out the first task, having 20 minutes to complete it. The tasks were a grid-based path optimisation problem requiring dynamic programming to compute the best possible resource accumulation under movement constraints (right/down or bidirectional) while handling cell-specific penalties/rewards (see tasks in the replication package \cite{martinez_montes_cristina_2025_15497559}). Upon completing the task or reaching the time limit, they completed a second survey (NASA TLX and SSSQ-post) reflecting on Task 1. The participants had a 10-minute break before moving on to the next task. After the break, participants watched a two-minute relaxation video, and right after, they started with Task 2, having 15 minutes to complete it. Both tasks were similar in complexity; however, Task 2 featured a shorter time limit to induce time pressure and increase stress levels. As with the first task, participants completed a survey evaluating Task 2. The experiment concluded with a final two-minute relaxation video to help the participants decompress.

\textbf{Exit Interview.} After participants took off the sensors, we ran a short interview to elicit their overall experience during the experiment (see interview in data collection methods). Finally, participants received a voucher for a restaurant to thank them for participating.

\subsection{Data Collection Methods}

In our study, we use a combination of biometric sensors and surveys to measure stress levels and mental workload. 

\subsubsection{Biometric Sensors}
\begin{figure}[htb]
    \centering
    \includegraphics[width=0.7\linewidth]{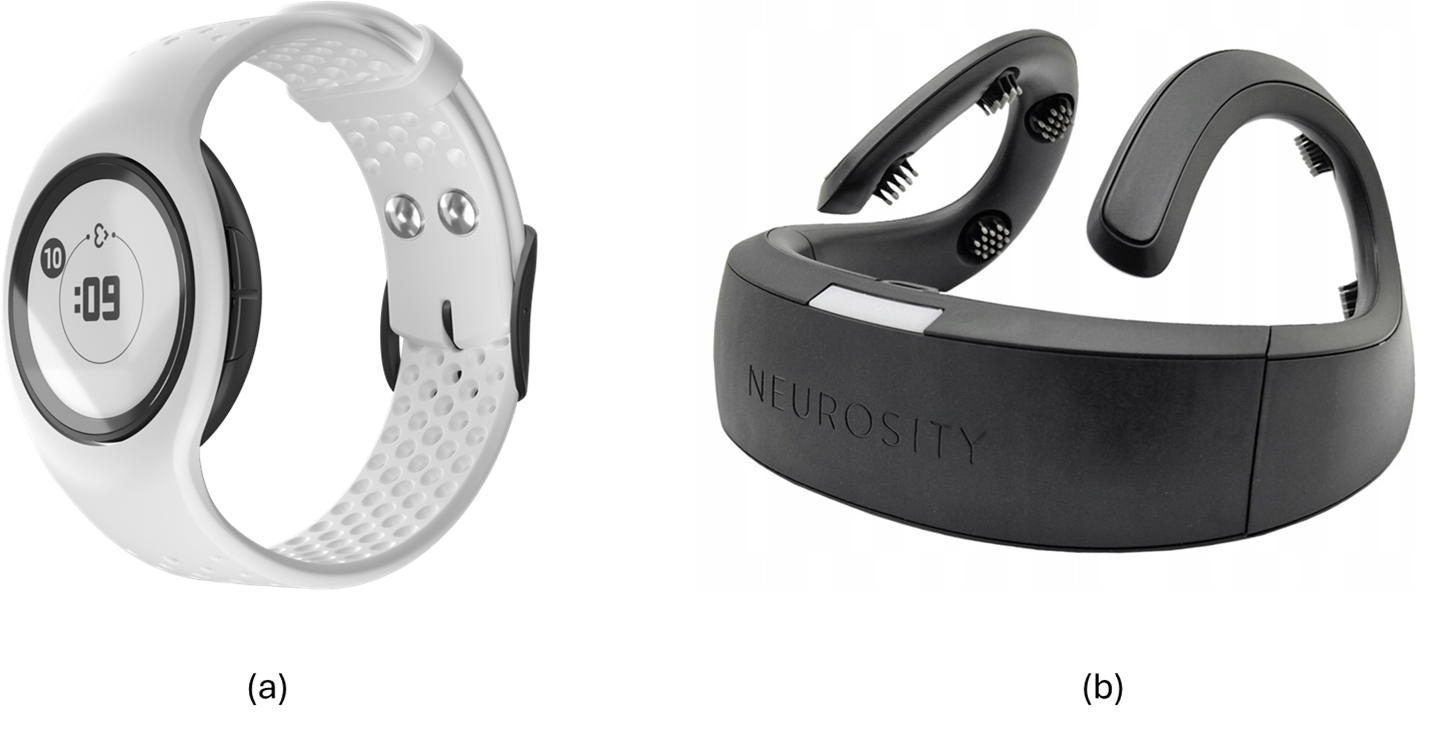}
    \caption{Wearable devices used in the study. (a) Embrace Plus by Empatica, (b) Neurosity Crown}
    \label{fig:instruments}
\end{figure}

To collect data, we utilised two biometric devices: a wearable wristband for EDA and HRV acquisition and an EEG helmet. The Empatica EmbracePlus\footnote{https://www.empatica.com/en-eu/embraceplus/}, as shown in Fig \ref{fig:instruments} (a), is 
a medical-grade wearable wristband, which we used for continuous, unobtrusive measurement of physiological signals. 
It includes a ventral EDA sensor that samples at 4 Hz and a PPG (photoplethysmography) sensor sampling at 64 Hz, from which we derived HRV metrics.
%pulse rate, EDA, skin temperature, and accelerometry-based activity metrics.
%This device features a ventral EDA sensor and a four-channel multi-wavelength photoplethysmogram sensor, providing a comprehensive approach to physiological data collection.
The EEG data were recorded using a Neurosity Crown device\footnote{https://neurosity.co/}, which measures electrical brain activity through its embedded sensors, as illustrated in Fig. \ref{fig:instruments} (b).
The device consists of eight channels (CP3, C3, F5, PO3, PO4, F6, C4, CP4, which acquired the brain signals at a sampling rate of 256 Hz. %The device's operating system applies pre-filters, including a 50–60 Hz notch filter with a 1 Hz bandwidth and a 2–45 Hz bandpass filter, both leveraging the Butterworth characteristic.

\subsubsection{Self-report Instruments}
We used a combination of consolidated self-reported instruments, which are explained in detail below and are widely adopted in the literature.

\textbf{Perceived Stress Scale (PSS-10)} \cite{lee2012review} is a widely used instrument to assess the degree to which individuals perceive situations in their lives as stressful. It evaluates feelings and thoughts over the past month, providing an understanding of how circumstances influence perceived stress levels.

\textbf{Short Stress State Questionnaire (SSSQ)} \cite{helton2015short} is a validated instrument to assess stress states, measuring the engagement of tasks, stress, and worry. It has demonstrated sensitivity to task stressors, with different task conditions producing distinct stress patterns consistent with prior predictions. The tool includes pre- and post-task versions, making it valuable for researchers studying conscious appraisals of task-related stress.
%\textbf{Self-Assessment Manikin (SAM)}~\cite{bradley1994measuring} is a non-verbal visual instrument that measures an individual's emotions and affective responses to an object or event.

\textbf{NASA Task Load Index (NASA TLX)}~\cite{hart2006nasa} was developed by the Human Performance Group at NASA's Ames Research Center, is a widely used tool for assessing subjective mental workload (MWL) during task performance. It evaluates MWL across six dimensions to produce an overall workload score: mental demand (cognitive effort for thinking, decision-making, or calculations), physical demand (intensity of physical activity required), temporal demand (time pressure involved), effort (exertion needed to maintain performance), performance (effectiveness in task completion), and frustration level (feelings of insecurity, discouragement, or contentment).

\subsubsection{Post-Task Interview}

The interview questions aimed to explore participants’ subjective experiences during the study, focusing on their stress levels and task-related perceptions. Participants were asked to describe their overall experience, including any factors contributing to their stress, and to reflect on specific moments of increased or decreased stress during the tasks. The questions also addressed the impact of wearing EEG and EDA devices on their concentration and performance. In addition, participants were encouraged to share any strategies they used to manage stress or maintain focus and were invited to provide further comments about their experience. See the questions in Table \ref{tab:interview_questions}

\begin{table}[h]
\centering
\caption{Interview Questions}
\label{tab:interview_questions}
\begin{tabular}{l p{10cm}}
\toprule
\textbf{No.} & \textbf{Question} \\
\midrule
1 & Can you describe your overall experience during the study? Did anything about the task or process contribute to your stress levels? \\
\midrule
2 & How did you feel during the tasks? Were there specific moments when you noticed increased or decreased stress levels? \\
\midrule
3 & How did you find the experience of wearing the EEG and EDA devices while completing the task? Did they interfere with your ability to concentrate or perform? \\
\midrule
4 & Did you use any strategies to manage your stress or stay focused during the task? If yes, what were they? \\
\midrule
5 & Do you want to add anything else about your experience in the study? \\
\bottomrule
\end{tabular}
\end{table}

\subsection{Data Analysis}

The analysis of each dataset is explained in the following subsections. The goal of our analysis is twofold. First, comparing the psychometric, i.e. the self-reported stress,  between Task 1 and Task 2 enables us to verify that we successfully induced stress in the participants during the second task by giving them less time for performing the coding task. Second, by comparing the biometrics collected during Task 1 and Task 2 we aim at verifying if there is any significant pattern in the physiological responses that can be used as a proxy for the self-reported level of stress at the end of each coding task. We further complement this analysis by comparing the biometrics during the pre-task and Task 1.  

\subsubsection{Psychometrics}
The psychometric data were analysed using RStudio. First, we cleaned the data, inspected for missing values, standardised variable names, and reversed scale items when needed (based on the psychometric instruments guidelines). Then, we calculated descriptive statistics for each psychometric scale, including means and standard deviations. We used the unweighted average of all the scales to report descriptive statistics and to compare results across time points (see table \ref{tab:descriptivestats}). Later, we tested the normality of the distribution and then applied a paired t-test to assess significant differences.

%Finally, SSSQ score was normalised using Z-scores to compare with the biometric results. The scripts and output are documented in an R Markdown file available in our replication package \cite{}.

%t-tests and p-value 

\subsubsection{Biometrics}
\label{sec:biometric_analysis}

To align the physiological signals with the different phases of the experiment (e.g., baseline, pre-task, first task, and second task), participants manually marked the start and end of each phase using the EmbracePlus wristband's event-tagging feature. These timestamps were then used to segment and synchronise the recorded physiological data (such as heart rate, skin temperature, and movement) with the corresponding experimental phases for the analysis. 

To compare stress levels between the two tasks, we extracted the raw data from the two physiological sensors for a 1-minute window, starting 30 seconds before the end of each task. We chose 30 seconds for the data extraction to account for the possibility that there could be a time gap between the switching of the tasks. This approach is in line with validated practices \cite{chae2021relationship}
The raw data extracted from the two sensors were processed differently. The approaches to processing raw data from two different physiological sensors are presented below. 

\textbf{EDA.} To account for individual differences in EDA signals, we standardised the signals using z-score normalisation relative to the baseline signal collected while watching the relaxing video, following established methods adopted in related work \cite{girardi2020recognizing}. The data was then preprocessed using the NeuroKit2 package \footnote{https://neuropsychology.github.io/NeuroKit/functions/signal.html}. We applied a 1Hz low-pass Butterworth filter to remove high-frequency noise, as done by Taylor et al. \cite{taylor2015automatic}. 
Next, we decomposed the filtered EDA signals into tonic (Skin Conductance Level, SCL) and phasic (Skin Conductance Response, SCR) components using the cvxEDA algorithm \cite{greco2015cvxeda}. This preprocessing step is required to separate the slow-varying tonic components from rapid phasic responses, both of which are relevant for detecting stress-related responses \cite{nardelli2022comeda}. %\nicole{Daniela: Please, explain why this is necessary by explaining the differences in the tonic and phasic phases and by indicating which one is relevant to stress recognition.} 
From these  two components, we extracted statistical features such as minimum, maximum, mean, and standard deviation (see Table \ref{tab:feature}), in line with previous work \cite{stress_picard} %\nicole{Daniela, add references to previous work using these features}.

Since EDA peaks can be interpreted as a response to stress episodes \cite{westerink2020deriving}, in our analysis we include consideration of such peaks. In particular, we identified EDA peaks using NeuroKit2's, following the peak detection approach proposed by Kim et al. \cite{kim2004emotion}. To account for variations in task duration, we computed the number of peaks per minute by dividing the total peak count by the duration (in minutes) of each experimental phase. Similarly, we consider the duration of the pre-task step for which we also extracted the biometrics. This normalisation allows for more reliable comparisons across conditions of different durations, as in our case.

\textbf{HRV.} Heart rate variability was calculated using the hrvanalysis library\footnote{https://github.com/Aura-healthcare/hrv-analysis}, based on interbeat intervals, which represent the time intervals between successive heartbeats. The signal was preprocessed by removing outliers (interbeat outside the 300–2000 ms range), as recommended by \cite{gulati2010heart}. Missing values were linearly interpolated, and ectopic beats were corrected using the Malik method \cite{malik1990heart}.

For each task, we computed the RMSSD (Root Mean Square of Successive Differences), which reflects short-term heart rate variability and typically decreases under stress \cite{yu2024exploring}. We also calculated the SDNN (Standard Deviation of Normal-to-Normal intervals), which has been shown to increase during stress episodes \cite{castaldo2015acute}. Finally, we computed the LF/HF ratio—the ratio of low-frequency (0.04–0.15 Hz) to high-frequency (0.15–0.4 Hz) components of the HRV power spectrum—which significantly increases during stress \cite{castaldo2015acute}.

\textbf{EEG.} Initial inspection of the raw EEG signals revealed several instances of missing data. Specifically, in two cases (P5 and P10), the first task was missing, and in three cases (P8, P9, and P10), the second task was missing due to technical issues during data acquisition. Additionally, data for the pre-task phase were missing for two participants (P6 and P10). As a result, we had 6 data points available for analysing the comparison between the first and second tasks, and 7 data points for analysing the comparison between the pre-task and first task. Therefore, we decided not to perform any statistical analysis.

\begin{table}[h]
\centering
\begin{tabular}{l|c|c}
\textbf{Task} & \textbf{EDA + HRV} & \textbf{EEG} \\
\hline
Filling Presurvey & 10 & 8 \\
First Task & 10 & 8 \\
Second Task & 10 & 7  \\
\end{tabular}
\caption{Number of datapoint per task (EDA, HRV and EEG)}
\label{tab:task_counts}
\end{table}

\begin{table}[ht]
\centering
\resizebox{\textwidth}{!}{
\begin{tabular}{lllll}
\toprule
\textbf{Modality} & \textbf{Type} & \textbf{Feature} & \textbf{Stress} & \textbf{Mental Workload} \\
\midrule
\multirow{4}{*}{EDA} & Tonic & Mean, std, min, max & $\uparrow$ \cite{reinhardt2012salivary} & - \\
\cmidrule(lr){2-5}
 & Phasic & Mean, std, min, max & $\uparrow$ \cite{yu2024exploring} & - \\
\cmidrule(lr){2-5}
 & EDA Phasic Peaks &  Count & $\uparrow$ \cite{westerink2020deriving} & - \\
 & EDA Raw Peaks &  Count& $\uparrow$ \cite{westerink2020deriving} & - \\
  & EDA Raw Phasic per Minute & Count & $\uparrow$ \cite{westerink2020deriving} & - \\
   & EDA Raw Peaks per Minute  & Count & $\uparrow$ \cite{westerink2020deriving}& - \\
\midrule
\multirow{3}{*}{HRV} & Time domain & RMSSD & $\downarrow$ \cite{yu2024exploring} & $\downarrow$ \cite{delliaux2019mental} \\
 &  & SDNN & $\uparrow$ \cite{castaldo2015acute} & - \\
\cmidrule(lr){2-5}
 & Frequency domain & LF/HF Ratio & $\uparrow$ \cite{castaldo2015acute} & $\uparrow$ \cite{delliaux2019mental} \\
\bottomrule
\end{tabular}}
\caption{Physiological Features: EDA and HRV with Expected Behavior under Stress and Mental Workload}
\label{tab:feature}
\end{table}

\textbf{Analysis.} We performed statistical analyses on EDA‑derived and HRV-derived features to compare performance between the first and second tasks. Each feature was first tested for normality using the Shapiro–Wilk test. When the normality assumption held, we used paired t‑tests; when it was violated, we substituted the non‑parametric Wilcoxon signed‑rank test.

\subsubsection{Interviews}
The interviews were analysed following the six steps of reflexive thematic analysis by Braun and Clarke \cite{braun2021thematic}. The interviews were first transcribed verbatim, the transcripts were read multiple times for familiarisation, and semantic, inductive codes were generated across the dataset.  Codes were then grouped to identify potential themes, which were reviewed and refined to ensure they accurately captured patterns in the data. Themes were clearly defined and named to reflect their core meaning, and selected quotes were used to illustrate each theme in the final write-up. The process was conducted manually, with careful attention to researcher reflexivity. Since we strictly followed Braun and Clarke guidelines and aligned with Big Q qualitative values, we did not assess for inter-coder reliability \cite[p. 240]{braun2021thematic} as coding was treated as a flexible, interpretative process.

\section{Results} \label{sec:results}

This section presents the biometrics, psychometrics and interviews (thematic analysis) data results. In particular, we report empirical evidence from the analysis of the psychometric and biometric indicators. 
%The relationship of the three types of data is detailed in the discussion section.

\subsection{Psychometrics}

Table \ref{tab:descriptivestats} summarises the descriptive statistics (mean, standard deviation, minimum, and maximum) of the psychometric instruments (PSS-10, SSSQ and NASA-TLX) for each measurement: Pre-task, Task 1, and Task 2. 

The average baseline perceived stress level (PSS-10) was in the moderate range (M=1.87, SD=0.56). Aligned with this, the pre-task SSSQ score (M = 2.35, SD = 0.38) also indicated a moderate subjective stress state before tasks began. Regarding their emotions, participants had a mean of 5.50 (SD = 1.10), reflecting relatively positive affect and moderate arousal levels before the tasks. 

After Task 1, SSSQ dropped slightly to 2.25 (SD = 0.50); this change was not significant enough to impact the group stress levels. Finally, the NASA-TLX results were also moderated (M = 9.83, SD = 3.60) with considerable variability across participants. 
Regarding Task 2, SSSQ-post's score remained relatively stable (M = 2.24, SD = 0.33). NASA-TLX increased slightly (M = 10.82, SD = 2.81); however, there was no significant change.

\begin{table}[htb]
\centering
\caption{Descriptive statistics per group and task}
\label{tab:descriptivestats}
\begin{tabular}{llcccc}
\toprule
\textbf{Group} & \textbf{Instrument} & \textbf{Mean} & \textbf{SD} & \textbf{Min} & \textbf{Max} \\
\midrule
\multirow{3}{*}{Pre-tasks} 
 & PSS-10 & 1.87 & 0.56 & 1.10 & 3.00 \\
 & SSSQ-pre   & 2.35 & 0.38 & 1.88 & 3.25 \\
 
\midrule
\multirow{3}{*}{Task 1} 
 & SSSQ-post   & 2.25 & 0.50 & 1.71 & 3.13 \\
 
 & NASA   & 9.83 & 3.60 & 3.17 & 15.00 \\
\midrule
\multirow{3}{*}{Task 2} 
 & SSSQ-post   & 2.24 & 0.33 & 1.79 & 2.71 \\
 
 & NASA   & 10.82 & 2.81 & 7.17 & 15.67 \\
\bottomrule
\end{tabular}
\end{table}

We compared participants' stress levels before and after the tasks using a paired t-test. Table \ref{tab:stress_tasks} presents the results of the two comparisons, showing that none of the comparisons were significant. For Pre-task vs Task 1, the t-value was 0.592, with a p-value of 0.569, indicating no significant change in stress from the pre-task phase to Task 1. Concerning Task 1 vs Task 2, results showed a t-value of 0.137 and a p-value of 0.893, which means no statistically significant difference in stress scores between Task 1 and Task 2. Participants reported comparable levels of stress during both tasks, which suggests that the reduced time for Task 2 was not enough to induce a stress condition during the second coding task.

\begin{table}[h!]
\centering
\caption{Comparison of stress scores between tasks }
\label{tab:stress_tasks}
\begin{tabular}{lccccl}
\toprule
\textbf{Comparison} & \textbf{t-value} & \textbf{p-value} & \textbf{Mean Difference} & \textbf{Interpretation} \\
\midrule
Stress: Task 1 vs Task 2 & 0.137 & 0.893 & 0.017 & No significant difference \\
Stress: Pre-task vs Task 1 & 0.592 & 0.569 & 0.092 & No significant difference \\
\bottomrule
\end{tabular}
\end{table}

Furthermore, we also tested for significant differences in NASA-TLX results, presented in Table \ref{tab:nasa_tasks}. Results showed a p-value of 0.426, implying there are no significant differences in mental workload in Task 1 and Task 2.

\begin{table}[h!]
\centering
\caption{Comparison of NASA-TLX scores between tasks}
\label{tab:nasa_tasks}
\begin{tabular}{lccccl}
\toprule
\textbf{Comparison} & \textbf{t-value} & \textbf{p-value} & \textbf{Mean Difference} & \textbf{Interpretation} \\
\midrule
 Task\_1 vs Task\_2 & -0.833 & 0.4265 & -0.983 & No significant difference \\
\bottomrule
\end{tabular}
\end{table}

% \begin{figure}
%     \centering
%     \includegraphics[width=0.6\linewidth, angle=-90]{images/SAM_group.pdf}
%     \caption{SAM Results per Group}
%     \label{fig:sam}
% \end{figure}

% Results from SAM are presented in Fig \ref{fig:sam}. We divided the scale into valence, arousal, and dominance to examine the findings closely. \nicole {We never mention Dominance in the background. If we decide to keep the results about the SAM questionnaire, we need to define also this third dimension of affect}. 
% Valence showed a mild but noticeable decrease in Tasks 1 and 2 in comparison with the baseline (Pre-task), which could suggest participants felt less happy or pleasant during both tasks.
% The changes in arousal were not significant either. Participants presented a decrease during task 1, which implies a feeling of reduced activation or alertness, possibly boredom or calmness. Regarding task 2, there was a slight increase; however, it was not big enough to reach the baseline. Overall, results showed that tasks were not highly stimulating.
% A similar situation happened with dominance. Participants initially felt very in control, but this decreased during Task 1 and partially rebounded during Task 2.
% Participants' emotions moved toward a less positive, less aroused, and slightly less dominant state during tasks than the pre-task state.

\subsubsection{Closer Look per Participant}

\begin{figure}
    \centering
    \includegraphics[width=0.7\linewidth, angle=-90]{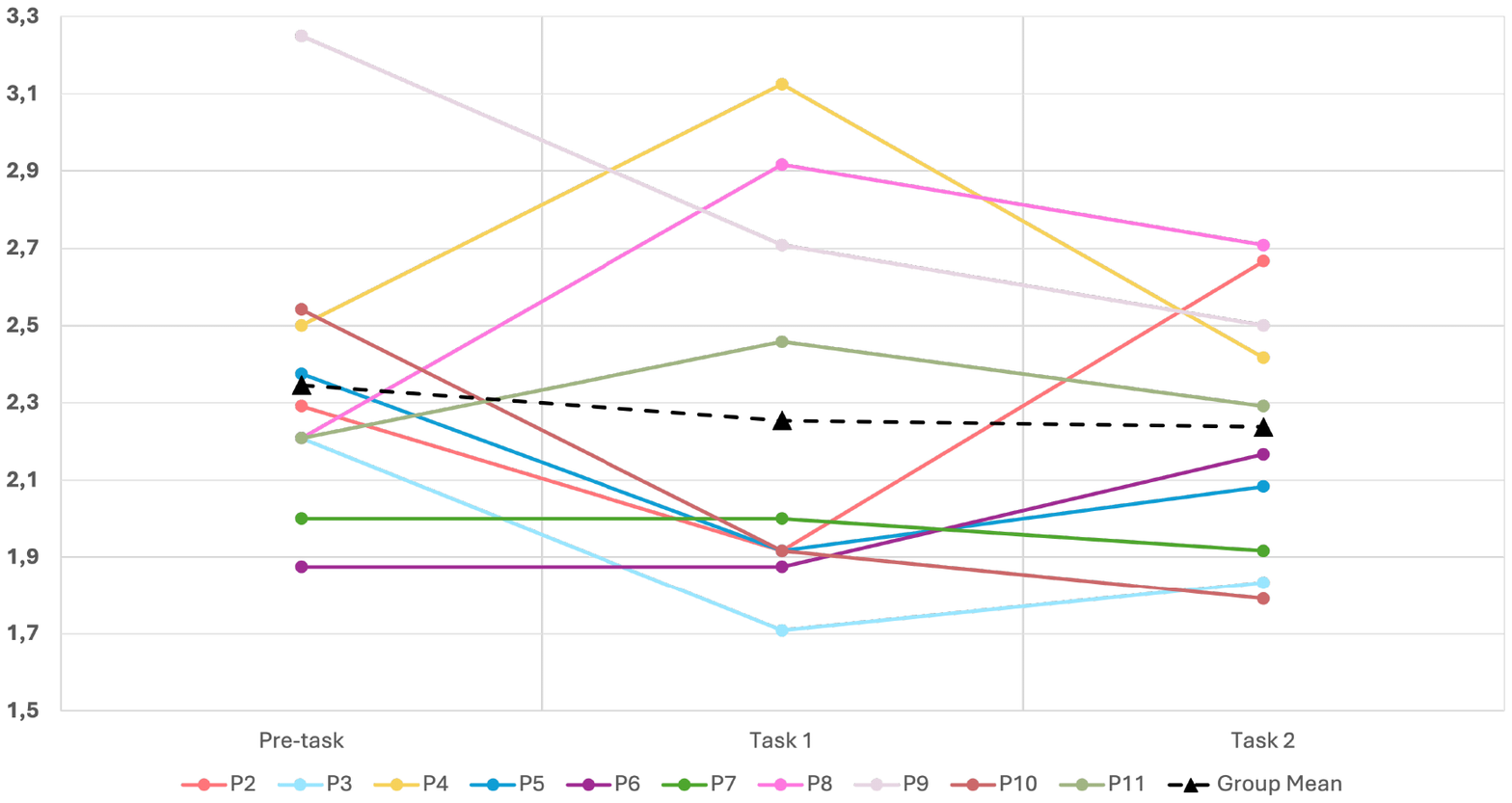}
    \caption{SSSQ Results per Participant. The Black line represents the group's mean.}
    \label{fig:stress_participant}
\end{figure}

\begin{figure}[h!]
    \centering
    \includegraphics[width=0.9\linewidth ]{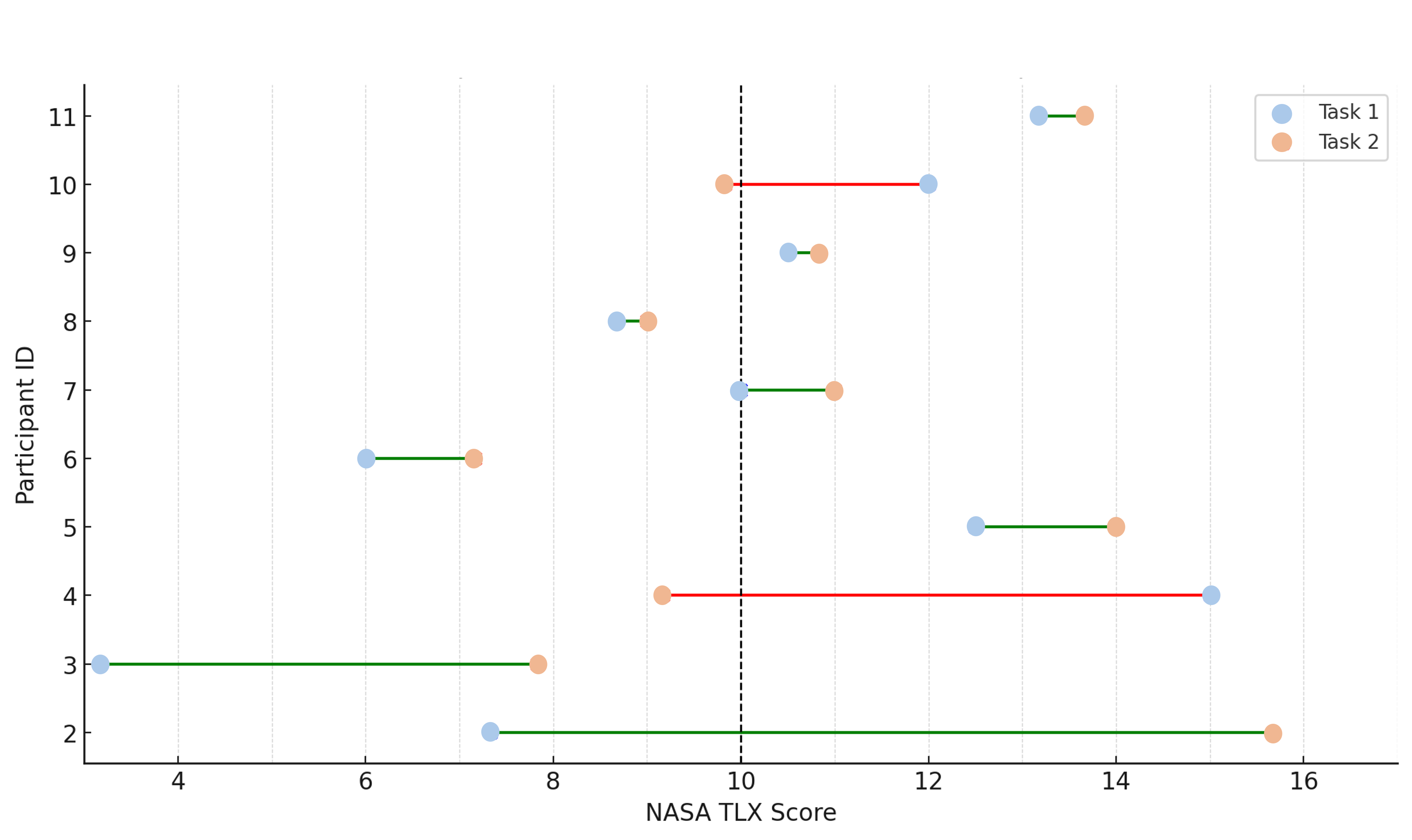}
    \caption{NASA-TLX results per participant. The image compares Task 1 and Task 2 scores. Green lines indicate an increase in score from Task 1 to Task 2, while red lines indicate a decrease. The vertical dashed line at score 10 represents the midpoint between low and high perceived workload.}
    \label{fig:nasa_participants}
\end{figure}

%\nicole{Cristy, why do we separate SSSQ from the previous subsection? Isn't this related to the part were we talk about stress?} 
We looked into each participant's stress levels before and after each task to better understand the low scores. Individual SSSQ results are shown in Figure \ref{fig:stress_participant}.

Participants generally did not show a pattern in their stress results; they had varied trajectories from their baseline to both tasks. 
For example, P3, P5, P9, and P10 decreased stress with tasks, possibly due to familiarisation or engagement. On the contrary, P4, P8, and P11 increased their stress, and P2 had an interesting trajectory with an initial decrease and finalising with a higher score than the baseline. Finally, P6 and P7 remained generally flat, showing stable stress during the experiment.

The variation in the responses reflects more the influence of individual differences and the task-specific experiences of participants than the tasks and the stressors we tried to add to the whole experiment.

%%%%NASA

Figure \ref{fig:nasa_participants} shows NASA-TLX scores per participant. The score ranges from 0 to 20, with 10 as the midpoint between low and high workload; higher scores indicate greater mental workload.
Most participants reported moderate to high workload in both tasks, with many scores clustering around or above the midpoint (10). Participants 2, 3, 5, 6 and 7 increased from Task 1 to Task 2 (green lines), which suggests Task 2 was more demanding for them. On the contrary, participants 4, 10 and 11 reported a decrease (red lines), indicating that they found Task 2 less demanding than Task 1. Two participants, P8 and P9, presented a minimal change in their scores from Task 1 to Task 2. Overall, the scores imply that Task 2 was perceived as more demanding by most participants, but responses varied considerably.

\subsection{Biometrics}

Our analyses of biometrics aimed at verifying if there are any statistically significant differences in the 15 % \nicole{Daniela, add the number of features overall} 
metrics features we extracted for Task 1 vs. Task 2, and Pre-task condition vs. Task 1. 
The results of our paired statistical tests revealed that the participants substantially exhibit the same behaviour between the conditions, with only one variable showing significant changes across different experimental conditions (see Table \ref{tab:results}). 

\begin{figure}[htb]
    \centering
    \includegraphics[width=0.7\linewidth]{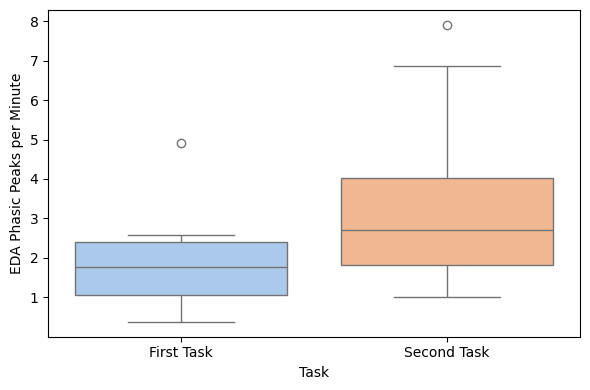}
    \caption{Differences between the EDA phasic peaks in the First Task vs. Second Task}
    \label{fig:peaks_first_second}
\end{figure}

In particular, when comparing Task 1 with Task 2, we observed 
%a significant decrease in relative gamma power at F6 ($t = 3.24$, $p = 0.02$, $n = 6$, paired t-test), as illustrated in Fig. \ref{fig:F6}. According to the literature \citep{minguillon2016stress}, an increase in relative gamma power is associated with higher stress conditions. As such, this pattern suggests an overall decrease of stress in Task 2, thus confirming the inability to induce a higher level of stress through time pressure in Task 2, as already observed in the analysis of the self-reported stress (see Section \ref{previous section here}). Additionally, we observed 
a significant increase in the EDA phasic peaks per minute between the first and second tasks (Wilcoxon signed-rank test: $W = 4$, $p = 0.01$, $n = 10$), indicating a higher increase in stress during the second task performance \cite{westerink2020deriving}, as shown in Fig. \ref{fig:peaks_first_second}. Furthermore, we observed a statistically significant difference in EDA peaks also between the pre-task baseline and the first task. This empirical evidence is in contrast with the self-reported stress, for which we did not observe statistically significant differences across the various experimental conditions.  
%Relative gamma power at C4 significantly increased from pre-survey to the first task (Wilcoxon signed-rank test: $W = 0$, $p = 0.02$, $n = 7$), as depicted in Fig. \ref{fig:C4}, thus suggesting that stress increases from the pre-tasks to the coding condition. Again, contrasting findings are observed for electrodermal activity, we found a highly significant decrease in phasic peaks per minute from pre-survey to the first task ($t = -4.5$, $p < 0.001$, $n = 10$, paired t-test) (see Fig. \ref{fig:peaks_pre_first}).
%No other features showed significant differences between the first and second tasks (all $p > 0.05$). However, due to the relatively small sample size, particularly for some EEG metrics where $n = 6$ or $n = 7$ due to data quality issues, these results should be interpreted with caution, as they may not generalize beyond the current dataset

%\cristy{I think it's important to add the results in numbers here}

%\begin{figure}[htb]
%    \centering
%    \includegraphics[width=0.8\linewidth]{images/relative_gamma_f6.png}
%    \caption{Relative Gamma Power (F6 Electrode) for Individual Subjects in Task1 and Task2}
%    \label{fig:F6}
%\end{figure}

%\begin{figure}[htb]
%    \centering
%    \includegraphics[width=0.8\linewidth]{images/relative_gamma_c4.png}
%    \caption{Relative Gamma Power (C4 Electrode) for Individual Subjects in Pre-task and Task1}
%    \label{fig:C4}
%\end{figure}

\begin{table}[ht]
\centering
\resizebox{\textwidth}{!}{
\begin{tabular}{lllll}
\toprule
\textbf{Comparison} & \textbf{Metric} & \textbf{Statistic} & \textbf{N} & \textbf{p-value} \\
\midrule
%First vs Second Task & relative gamma (F6) & T = 3.24 & 6 & 0.02 \\
First vs Second Task & eda phasic peaks per minute & W = 4 & 10 & 0.01 \\
%Pre-task vs First Task & relative gamma (C4) & W = 0 & 7 & 0.02 \\
Pre-task vs First Task & eda phasic peaks per minute & T=-4.5 & 10 & 0.00 \\
\bottomrule
\end{tabular}
}
\caption{Statistical results for comparisons between tasks.}
\label{tab:results}
\end{table}

\begin{figure}[ht]
    \centering
    \includegraphics[width=0.7\linewidth]{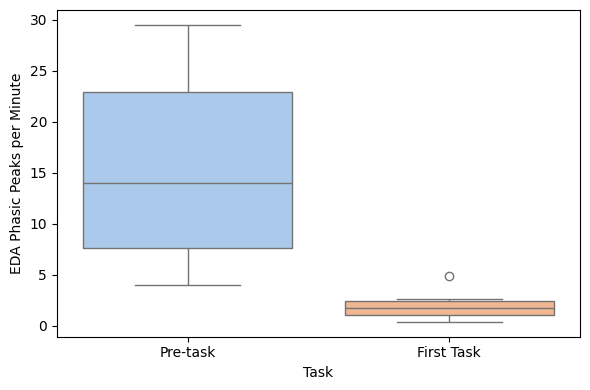}
    \caption{Differences between the EDA phasic peaks in the Pre-task vs. First Task}
    \label{fig:peaks_pre_first}
\end{figure}

We interpreted this contrasting finding as an indication of two possible problems: (i) the misalignment of self-reported and actual stress of participants, with SSSQ and biometrics indicating opposite findings; (ii) a high degree of diversity in the stress experienced by participants, as also suggested by Figure~\ref{fig:stress_participant}. 
To obtain deeper insights and in search of an explanation for this mixed evidence, we conducted a follow-up analysis to verify the alignment between the self-reported stress and the EDA peaks, which we describe in Section~\ref{sec:alignement}.

\subsection{Thematic Analysis}

Four themes were generated from the interviews with the ten participants. Below we elaborate on each of them and their corresponding sub-themes. 

\subsubsection{\textbf{Theme 1: Task Impressions: Engagement and Learning}}

This theme describes several aspects of the experience, including engagement and skill development. Despite the challenges, many participants found the tasks engaging and appreciated the opportunity to learn and apply their skills. The theme also elaborates on the participants' perception and experience of programming while wearing devices to collect their biometrics.

\textit{\textbf{Sub-theme 1: Perceived Task Structure and Difficulty}}

This sub-theme captures how participants perceived the tasks, including their clarity, complexity, and how their impressions evolved. In general, perception varied mainly regarding time constraints and clarity. Some found the second task straightforward, with no significant obstacles. As one participant stated: 

\textit{``Everything was clear, so I didn't have any problems during the second task. Maybe I was feeling that the time was less, but just a little feeling, but nothing else".} P6

The quote suggests that the task was well-structured and manageable for some despite the tighter time frame.
However, other participants shared their difficulties with time pressure, which influenced their confidence and overall experience. One participant described feeling confident that they would not be able to complete the second task within the given time, leading to a particular feeling: 

\textit{``In the second time in the second task, I was quite sure that I couldn't complete that in time, so I felt a bit unhappy about that".} P2

The quote is a good example of how time constraints, rather than the complexity of the task itself, shaped the perceptions of difficulty.
Additionally, some participants noticed structural similarities between the two tasks, which helped them refine their approach to the second one. However, this familiarity did not always mitigate concerns about time limitations, as some questioned why a more difficult task was allotted less time. These differing perspectives suggest that task difficulty was not purely objective but influenced by individual expectations, time-related pressure, and the ability to adapt strategies based on prior experience.\\

\textit{\textbf{Sub-theme 2: Engagement and Learning Through Task Progression}}

This sub-theme focuses on how participants engaged with the tasks, their sense of accomplishment, and the learning process they experienced. Participants' engagement was often shaped by their initial impressions and their ability to adapt. Some participants initially underestimated the complexity of the first task but later recognised deeper patterns that made it more challenging. 

For example, Participant 10 reflected on their shift in perception, stating:

\textit{``When I read the first one at the beginning, I thought I had all figured it out, but then I read it again, and I saw some patterns that suggested that it was more difficult than I thought, and so maybe in that moment, I felt that it could be difficult and that I already knew that  I could not be able to solve it in 20 minutes".} P10

For Participant 10, engagement was not static but developed as they reassessed their approach and deepened their understanding of the task.
The experience of completing the first task also influenced how participants approached the second. Some found familiarity with the structure helped them engage and work through the task more efficiently. 

One participant described, \textit{``Maybe because after completing the first, I reasoned better about the task because they were somehow similar. And the second, it was easier for me to solve it, and it was fine".}   P5

The quote reflects how task progression supported learning, enabling participants to refine their problem-solving strategies and improve efficiency.
Overall, engagement was influenced by the task challenge and the opportunity to apply and adapt knowledge. While some participants encountered unexpected difficulties that affected their confidence, others found that progressing through the tasks enhanced their ability to approach problems more effectively.

\subsubsection{\textbf{Theme 2: Emotional Responses to Challenge and Uncertainty}}

This theme elaborates on the participants' emotional journey as they navigated the tasks. Among the various emotions participants experienced were frustration, stress, and irritation, particularly when tasks were challenging or when they were uncertain about how to proceed. 
Time constraint was mentioned as a significant stressor for participants; for example, Participant 11 expressed: 

\textit{``The only thing that contributed to the stress level was the time. I feel like with the task, 20 minutes was a little like the first task and the second task. 20 and 15 minutes maybe were a little bit too little".}  	P11

Participant 11's quote is interesting because it dismisses other potential stressors like task complexity, unfamiliarity, or difficulty. It also shows how strongly time pressure could affect performance and well-being.
 
Other participants found the time limit particularly frustrating when encountering difficulties, such as recalling specific programming libraries or debugging errors. Similarly, participants noted that uncertainty about their solutions contributed to their stress, mainly when they could not test or verify their code.

Despite the challenges, some participants viewed stress as a natural part of problem-solving, accepting moments of frustration as inherent to the coding process.

\subsubsection{\textbf{Theme 3: Sources of Distraction and Discomfort}}

Some factors influenced participants' attention, perception, and, in some cases, emotions. This theme explores how the experimental setup and environment impacted their experience.
Elements within the environment disrupted their focus or contributed to discomfort, affecting their ability to fully engage with the tasks.
The context of the experiment also played a role. For example, Participant 10 mentioned that they might have performed better if they had been alone. The presence of others seemed to increase their distraction, as they became more aware of how they were performing in comparison

\textit{``Maybe I would have performed better if I was alone in the room, because maybe I would have started talking by myself and so on".} P10

Internal distractions, such as self-conscious thoughts about performance and concerns about how others were doing, were also noted. For example, Participant 4 mentioned:

\textit{``I have some thoughts about others or my or my results, I try to stay focused and I like, like, try to push away the thoughts and concentrate (like how I'm performing, if others are performing, well, )".}   P4

This participant tried to push these thoughts aside to stay focused, but they remained a persistent challenge. 

These distractions and discomforts added to participants' challenges in maintaining attention and emotional balance during the study.

\subsubsection{\textbf{Theme 4: Coping Strategies and Adaptation}}

Participants employed various strategies to manage stress and maintain focus. This theme explains how participants adapted to the challenges of the experiment by using these strategies. They used task decomposition,  deliberate attentional control, and actively ignoring feelings and physical actions. 

Some participants distanced themselves from the emotional weight of the task by reminding themselves that the experiment was not an exam, thereby reducing performance pressure. Others reported ignoring negative feelings entirely and concentrating on solving the problem instead. Using a structured approach to problem-solving was also mentioned; a specific example is Participant 3, as expressed in the following quote: 

\textit{``I was thinking about the best way to approach it, like if I should start by defining functions, because that's what I usually do or not, go straight forward to the code without anything in any method at all".}	  P3

P3 decision-making process to approach the task seems to rely on prior experience and habitual strategies, notable in the phrase ``because that's what I usually do". This participant showed flexibility in adapting their approach based on the task's demands. Task planning and adaptation add additional mental workload to the tasks, which could also impact our quantitative results. Overall, these adaptive behaviours allowed participants to mitigate stress and maintain productivity within the experimental setting.

\section{Discussion} \label{sec:discussion}

In this section, we discuss how we answered our RQs, the key contributions of this study and the threats to validity.

While we aimed to assess the alignment of psychometric instruments and biometric data and find stress patterns in these data sources, our results did not offer consistent evidence to support clear conclusions. Nonetheless, our results offer indicative insights that may inform future research directions.

To answer our RQs:

\textbf{RQ1: How reliable are psychometric stress measures compared to real-time biometric indicators (EEG and EDA) during software engineering tasks?}

Psychometric results showed, at a general level, moderate stress. Furthermore, there was no increase in stress from pre-tas to Task 1 nor from Task 1 to Task 2. Mental workload results were around the moderate level, too, and showed no significant differences from Task 1 to Task 2. Aligning these results with biometrics, for EDA, only one metric (phasic peaks per minute) showed a statistically significant difference across tasks. However, this single biometric indicator did not consistently align with the psychometric instruments. For EEG, we lost several data points. Hence, we could not analyse it, and we lost that comparison. 

Consequently, we cannot draw firm conclusions about the reliability of psychometric measures relative to biometric data. Hence, our findings are only indicative.

\textbf{RQ2: What stress-related patterns can be identified in real-time biometric data (EEG and EDA) during software engineering tasks?}

Since we could not analyse the EEG data, our observations were only on EDA metrics. Phasic peaks per minute were the only metric showing significant differences across tasks, which could suggest a stress-related pattern. However, our results are limited since we did not find any other trends in the rest of EDA  features. We do not have robust enough evidence to establish precise or generalisable patterns in this context.

\subsection{Main Contributions}

This study offers the following insights.

\subsubsection{\textbf{Evidence Supporting the Alignment of Biometrics and Psychometric Instruments}}
\label{sec:alignement}

One of our goals in this study was to find to what extent biometrics align with self-report. Furthermore, the mixed findings observed for the analysis of the biometrics and psychometric instruments call for further analysis of the alignment of the self-reported stress and the biometrics for each participant. 
In fact, as we report in the previous section, although our quantitative results of SSSQ showed no significant differences in stress levels, we observed variations of EDA peaks across the experimental condition and, in particular, between pre-task and Task 1 and between Task 1 and Task 2 that suggest the participants might have actually experienced some stress episodes. 

In search of an explanation, we performed a follow-up analysis by triangulating self-reported stress, EDA peaks and results of the qualitative coding of interviews. By applying a data triangulation across our data collection methods, we observed that our multi-modal measurement approach was sensitive to the same variations in stress responses. At the group level, there were no significant differences in stress, emotions, and mental workload levels between Tasks 1 and 2 in psychometric and biometric results. Specifically, we decided to look at each participant's behaviour with a focus on stress using self-reported stress, interviews, and EDA peaks.

In Table~\ref{tab:stress_followup}, we indicate the self-reported level of stress based on the answers participants provided for the PSS-10 (second column) and SSSQ (third, fourth, and fifth columns). We used the stress levels in the table, mapping the numerical answers to the corresponding levels of each psychometric instrument, for example, for PSS-10 ``Low, Moderate and High" stress \cite{adamson2020international}; and for SSSQ ``Not at all, Somewhat, Very much and Extremely" \cite{helton2015short}. We remind the reader that the PSS-10 reflects perceived stress over the past month, while the SSSQ captures stress levels pre- and post-tasks. Colour coding indicates stress intensity: green = low/no stress, orange = moderate stress, red = high stress.
Finally, in the last column, we report excerpts from the participants' answers during the interviews that pertain to the stress experience and the stress triggers they reported, if any. Interview quotes give contextual insight into individual experiences. 

\begin{figure}[htb]
    \centering
    \includegraphics[width=1\linewidth]{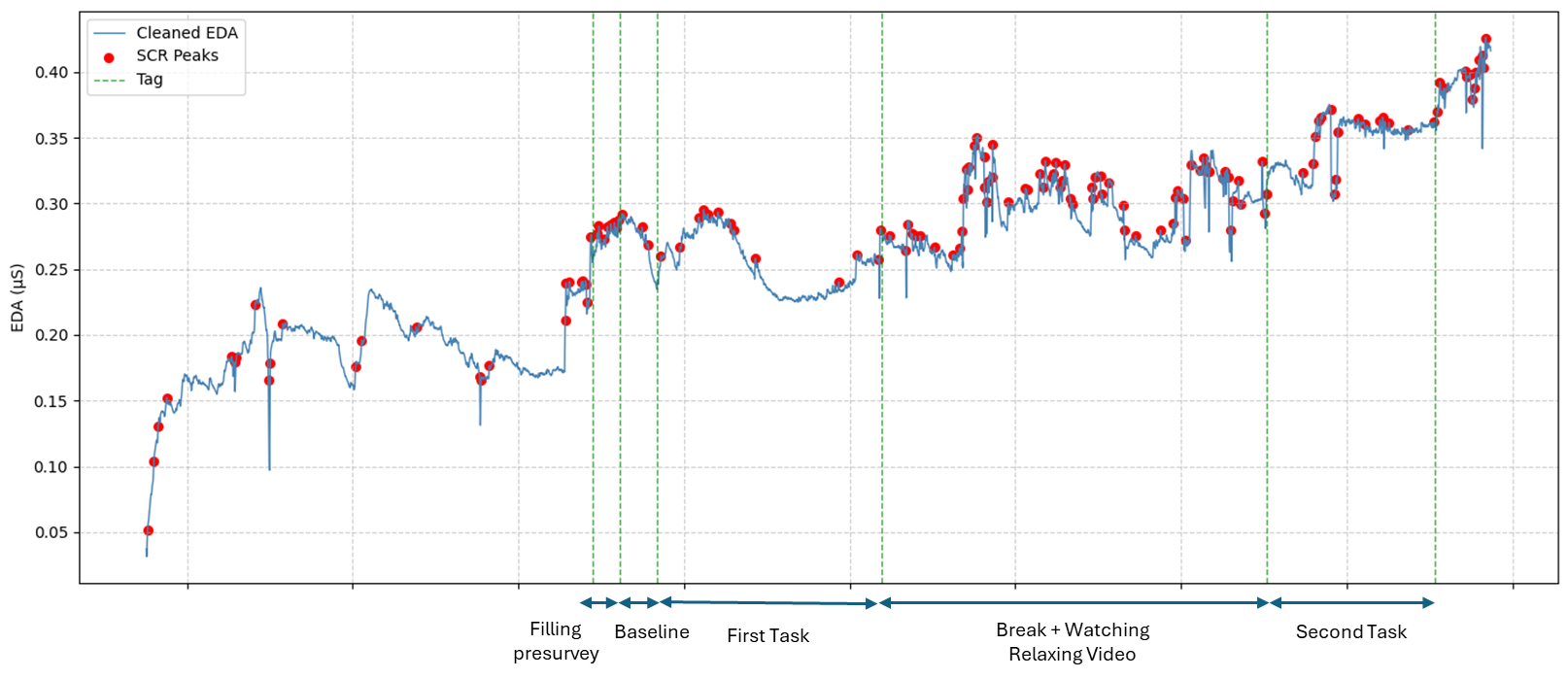}
    \caption{EDA signal and its peaks (red dots) across the experimental phases  for P9}
    \label{fig:EDA_P9}
\end{figure}

\begin{figure}[htb  ]
    \centering
    \includegraphics[width=1\linewidth]{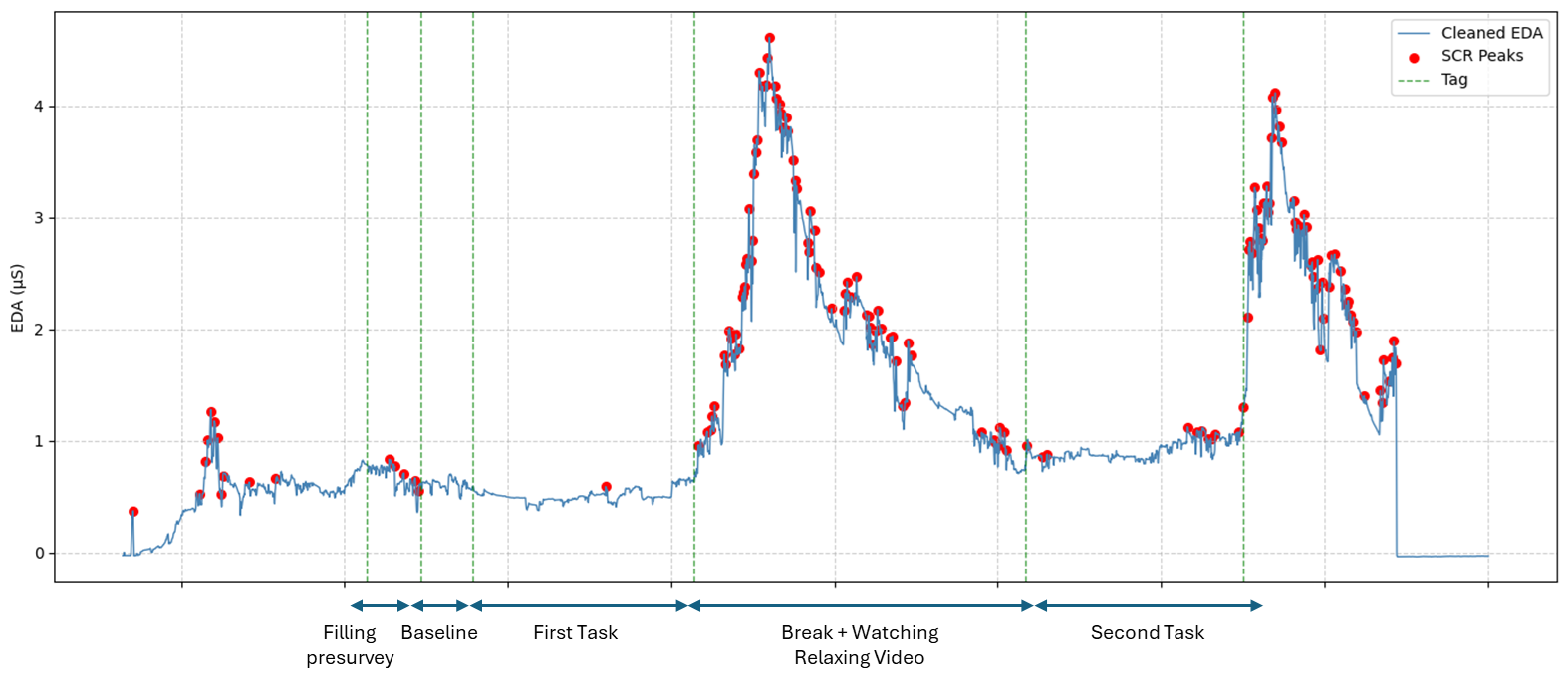}
    \caption{EDA signal and its peaks (red dots) across the experimental phases for P3}
    \label{fig:EDA_P3}
\end{figure}

To complement this multi-modal analysis, we triangulate the data in the table with plots of the EDA peaks (see Figures~\ref{fig:EDA_P3} and~\ref{fig:EDA_P9}). 
The green vertical lines correspond to tags created by participants during the task to contextualize events (e.g., beginning of pre-survey, beginning of baseline, start and end of first task, start and end of second task). The red dots highlight the peaks obtained following the same approach described in Section~\ref{sec:biometric_analysis}. 

\begin{table}[htb!]
\centering
\resizebox{\columnwidth}{!}{%
\begin{tabular}{p{0.2cm}p{1.2cm}p{1.2cm}p{1.2cm}p{1.2cm}p{6.5cm}}
\hline
\textbf{ID} & \textbf{PSS-10} & \multicolumn{3}{c}{\textbf{SSSQ}} & \textbf{Interviews} \\
 & & Pre task & Task 1 & Task 2 & \\
\hline
 \\
P2 & \textcolor{orange}{Moderate} & \textcolor{orange}{Moderate} & \textcolor{ForestGreen}{No stress} & \textcolor{orange}{Moderate} & ``I was okay during the first task, even though I had the sensation of being unable to solve it, it was more or less okay. The second task was more stressful, I think because it was time-pressured." \\
\hline
\\
P3 & \textcolor{orange}{Moderate} & \textcolor{orange}{Moderate} & \textcolor{ForestGreen}{No stress} & \textcolor{ForestGreen}{No stress} & ``It was kind of fine. I didn't really get stressed. The only part where I was a little stressed was during the code, but not significantly." \\
\hline
\\
P4 & \textcolor{orange}{Moderate} & \textcolor{orange}{Moderate} & \textcolor{orange}{Moderate} & \textcolor{orange}{Moderate} & ``I felt frustrated, stressed because I couldn't solve the problem, I didn’t understand what was wrong and that caused me some stress. I wanted to solve it but couldn't." \\
\hline
\\
P5 & \textcolor{orange}{Moderate} & \textcolor{orange}{Moderate} & \textcolor{ForestGreen}{No stress} & \textcolor{orange}{Moderate} & ``Overall, it was funny because I was curious about my stress level. During the second task I was more focused and that helped. The stress was moderate, nothing overwhelming." \\
\hline
\\
P6 & \textcolor{ForestGreen}{Low} & \textcolor{ForestGreen}{No stress} & \textcolor{ForestGreen}{No stress} & \textcolor{orange}{Moderate} & ``I was pretty okay during everything about the task. I didn’t feel particularly stressed, but I did notice a slight increase when the second task started, probably due to the pressure to complete it quickly." \\
\hline
\\
P7 & \textcolor{orange}{Moderate} & \textcolor{orange}{Moderate} & \textcolor{orange}{Moderate} & \textcolor{ForestGreen}{No stress} & ``When I understand what I do, my stress goes down. In the second task, once I figured out the logic, I felt relaxed and enjoyed finishing it." \\
\hline
\\
P8 & \textcolor{orange}{Moderate} & \textcolor{orange}{Moderate} & \textcolor{orange}{Moderate} & \textcolor{orange}{Moderate} & ``At the beginning, I didn't understand anything, which stressed me out. But after a while, I got into the flow and things became easier. It was challenging but not too difficult." \\
\hline
\\
P9 & \textcolor{red}{High} & \textcolor{orange}{Moderate} & \textcolor{orange}{Moderate} & \textcolor{orange}{Moderate} & ``I was not stressed at all in the first task because I found it quite simple. But in the second task, I got a bit stuck and it was a bit stressful. Still, I managed to finish." \\
\hline
\\
P10 & \textcolor{orange}{Moderate} & \textcolor{orange}{Moderate} & \textcolor{ForestGreen}{No stress} & \textcolor{ForestGreen}{No stress} & ``I didn't feel stressed because I didn't feel like I was being evaluated. It felt like an exercise more than a test, so I remained calm throughout." \\
\hline
\\
P11 & \textcolor{orange}{Moderate} & \textcolor{orange}{Moderate} & \textcolor{orange}{Moderate} & \textcolor{orange}{Moderate} & ``The only thing that contributed to the stress level was the time. I felt a bit pressured to complete it fast. That made me more focused but also raised the stress a bit." \\
\hline
\end{tabular}%
}
\caption{Alignment of psychometric stress measures (PSS-10 and SSSQ) with qualitative interview excerpts across participants.}
\label{tab:stress_followup}
\end{table}

%\begin{figure}[htb]
%    \centering
%    \includegraphics[width=1\linewidth]{images/P6.png}
%    \caption{EDA signal and its peaks (red dots) across the experimental phases  for P6}
%    \label{fig:EDA_P6}
%\end{figure}

Looking at Table~\ref{tab:stress_followup}, we observe that 4 out of 10 participants (P4, P8, P9, P11) have the same stress level (SSSQ) for the entire study, including the pre-task. This suggests that the experimental setting did not induce any changes in the self-reported stress compared to the pre-task condition. In fact, when we plot the EDA signal and its peaks for one of these participants (P9), we do not observe significant variations in the signal behaviour, with a slight increase towards the end of Task 2. This aligns with the self-report by P9, saying that '\textit{I was not stressed at all in the first task because I found it quite simple. But in the second task, I got a bit stuck, and it was a bit stressful. Still, I managed to finish}'. 
Moreover, for P9, Figure \ref{fig:EDA_P9} shows a gradual increase in tonic levels throughout the session, with frequent SCR peaks (marked in red). Regarding the pre-task, EDA is relatively low and stable, with a few SCRs, possibly due to the setting or anticipation. Later, in Task 1, EDA rises slightly but stabilises mid-task. Additionally, several SCR peaks are present but not densely clustered. Physiological arousal is moderate, consistent with SSSQ results. Finally, in Task 2, there is an increase in tonic EDA and dense clustering of SCR peaks, indicating possible heightened stress/arousal during this task. Psychometric results report Task 2 as "moderated" stress, with no significant changes from Task 1, as in the EDA results, and consistent with NASA-TLX (very close to the midpoint) results as well. Considering the interview quote, P9 reflects frustration and helplessness rather than classic stress, explaining why there were no changes in the SSSQ, but it is strongly visible in the EDA.

Similarly, P4, P8, and P11 report a mild experience of frustration or stress during the tasks but do not mention that they experienced different levels of stress across the conditions, which aligns with the consistent 'moderate' stress scores reported throughout the experiment. These alignments help contextualise psychometric scores. For example, participants P4 and P8 consistently reported moderate stress across the SSSQ, and their interview excerpts corroborate the presence of frustration and task-related cognitive effort. 

For participants P3 and P10, we observed that they were in a pre-task stress condition, while they were not stressed while coding. This could indicate that, for some participants, stress might be induced by the idea of participating in the study, with stress subsequently dropping down during the actual coding tasks. In fact, P3 reports that `\textit{It was kind of fine. I didn't really get stressed. The only part where I was a little stressed was during the code, but not significantly.}' When looking at P3's EDA plot, we can observe behaviour that aligns with the self-report, with peaks completely absent during both Task 1 and Task 2 (see Figure~\ref{fig:EDA_P3}). Adding the mental workload results to the analysis (see Picture \ref{fig:nasa_participants}), for P3, none of the tasks was challenging; they stayed on the low-demand side for both tasks. Similarly, for P10, even though they were above the median (moderate level), they did not reach a considerably high level of mental demand.

%A different behavior is observed for P6, who report a moderate stress in Task 2. Based on the explanation provided in the interview, this is due to the perception of the time-pressure in Task 2, for which the participants report a slight increase of stress at the beginning. 

%For participants P2, P8, P11 we see that they are not in stress during the pre-task assessment but are stressed during both coding task, This could be an indication of the fact that the shorter time for Task 2 does not induce higher stress but, overall, they are stressed during coding more than in their normal (non-conding) condition. Again, could you please check if they report anything related to this in their interviews?

%Furthermore, there is agreement on the data when biometrics are included in the triangulation. For example, Figures \ref{fig:EDA_P3} and \ref{fig:EDA_P6} shows P3 and P6's EDA results, respectively; there are valleys of reaction in pre-task, tasks 1 and 2, consistent with psychometric results (PSS-10: Low and SSSQ: No stress/moderate). 

Importantly, this alignment does not suggest perfect correlations but rather lends support to the credibility of our measurements. Research has shown that psychophysiological responses to stress are highly individual and context-dependent \cite{hellhammer2009salivary, thayer2012meta}, and subjective stress awareness may not always be linearly related to biometric signals. 

The observed consistency across data sources reinforces their general alignment and the value of this study of using a multi-method approach for stress detection in complex, cognitively demanding tasks like programming.

\paragraph{Difference of Acute and Long Term Stress}

There are several considerations to consider when measuring a specific type of stress in experiments. Our target was to measure acute stress, which at an emotional level, refers to the appraisal resulting from situations evaluated as threatening and overwhelming based on the individual's available coping resources~\cite{sandner2020investigating}. Reactions to this type of stress are complex and multidimensional; therefore, assessing its impact requires an equally nuanced approach \cite{sandner2020investigating}. We used a measurement approach to capture multiple aspects of acute stress. This included a baseline recording of EEG and EDA before the task, the PSS-10 to assess participants' general stress perception over the past month and the SSSQ pre-task. 

Despite these measures, the distinction between acute, subjective and long-term (chronic) stress remains a significant challenge. Biometric tools such as EEG and EDA are well-established for detecting acute stress markers. For instance, EDA reliably reflects sympathetic arousal \cite{critchley2002electrodermal}, while EEG patterns can identify emotions, vigilance, mental workload and stress levels \cite{hou2015eeg}.

However, longer-term stress exposure can influence biometric signals. Seo et al. \cite{seo2010stress} found that long and repetitive exposure to stress affects the ability to regulate cortisol levels. Further, since there are relationships between salivary cortisol levels and physiological variables (e.g. heart rate and galvanic skin response), chronic stress may alter baseline autonomic activity, mitigating or distorting the physiological changes typically associated with acute stress responses.

Therefore, even with a pre-task baseline, participants with high long-term stress may have exhibited attenuated or irregular acute stress responses during the task. For instance, a chronically stressed individual may be less responsive to our experimental stressor, leading to smaller physiological and psychometric shifts and potentially contributing to null findings. Moreover, our population in this study, PhD students, have a higher vulnerability to mental health difficulties compared to the general population, with multiple studies indicating elevated levels of anxiety, depression, and overall psychological distress in this group \cite{friedrich2023your}, which makes it even more possible the existence of long-term stress. These ongoing stress conditions could have added noise to the biometric and psychometric data. As a result, the biases introduced by chronic stress exposure may have masked clearer patterns of acute stress, limiting the sensitivity of our measures to detect short-term changes.

%How do the different types of stress impact the biometrics and psychometric results?
%Risks of mismeasure.
%Were participants already stressed?
%Biometrics might collect more of the long-term stress(?)

\subsubsection{\textbf{Methodological Insights on Recruitment and Experimental Design}}

We discussed several insights on the experimental design and implementation of the study and offer suggestions for improvement that we would apply in a future study.

\paragraph{Recruitment}
The participants in this study were mainly PhD students. As laid out in Sec.~\ref{sec:alignement}, PhD students tend to experience a high level of long-term stress and are more prone to stress-related mental health challenges, which may have influenced the results. We are hoping to replicate this study in an industrial practitioner context for further insights. \textit{Suggestion:} Recruit software developers in industry instead of PhD students.
 
\paragraph{Mental Workload and Stress} 
Both the biometric and the psychometric results show a lack of induced stress in the participants. However, while the participants did not show stress on biometric or psychometric scales, they did report stress in interviews. Hence, there is a psychological component of stress that does show up to some extent in the qualitative data, but less so in the quantitative data (only partially in the NASA-TLX results, see Fig.~\ref{fig:nasa_participants}, but not to a statistically significant extent), see Tab.~\ref{tab:nasa_tasks}. One possible explanation is that extra effort was exerted to meet the high demands of the task by mobilising extra energy through mental effort~\cite{gaillard1993comparing}. Since our EEG data measurements were insufficient, we cannot compare directly to the results of Mohanavelu et al.~\cite{mohanavelu2020dynamic} or Martínez Vásquez et al.~\cite{martinez2023mutual}. \textit{Suggestion:} Include additional instruments for differentiating mental workload from stress.
 
\paragraph{Distractions} One participant mentioned that the printer starting a job was distracting. Furthermore, that sometimes led other people to enter the room. While such distracting factors, e.g., other people in the room, take away from the setting of a controlled experiment, they can be linked to a more realistic setting than participants being in a room by themselves. Hence, it leads to a better representation of a real-world scenario. \textit{Suggestion:} Control for distractions in replication.

\paragraph{Participant Motivation} The participants' motivation was probably not strong enough.
Since participants did not have pressure to do this task well, as it had no consequences for them, this could be an indicator of why they did not get stressed as expected by the shortened time available to them for Task 2. \textit{Suggestion:} Pick a task that the participants care about and want to see through.

\paragraph{Increased Stimuli to Induce Stress}
Other ways of inducing stress, e.g., a simulated power outage, switching off the light, loud noise, or pretending this is their exam, would be unethical. However, if the reason the time pressure failed to induce stress is really due to a lack of motivation of the participants to succeed, then none of these are likely to make a difference.
\textit{Suggestion:} Introduce a stronger incentive through a higher remuneration if a task is completed. 

\paragraph{Technical Challenges of Sensors} 
The practical and technical feasibility of using biometric sensors in real-world or semi-controlled software engineering tasks is limited due to the fact that sensors are not the most robust or reliable. In combination with a limited time slot that the participants were booked for, this did not allow much room for error. If it was not detected immediately when a sensor was not working, we lost data on that participant. \textit{Suggestion:} Plan more buffer time.
% Why certain stimuli may not succeed in eliciting stress?
% Recommendations for future designs (e.g., stronger or ethically viable stressors, better recruitment criteria).
%Why the study failed to induce stress (e.g., weak motivation triggers, insufficiently stressful external factors like time pressure)
%Suggestions for improvement (e.g., stronger stressors like noise, interruptions, or ethical-but-more-effective triggers)

%Data quality issues and sensor usability problems ( we lost a lot of data).

%Lack of a pilot group.

\subsubsection{\textbf{General Lessons for Conducting Stress Studies}}

Our study, despite yielding negative results, shed light on methodological challenges in eliciting measurable stress responses within the context of software engineering experiments.

The reduction of task completion time from 20 to 15 minutes failed to elicit measurable stress responses, suggesting that time pressure alone may not serve as an adequate stressor for experienced programmers. As one participant remarked, "The only thing that contributed to the stress level was the time," indicating that while time constraints were perceived, their impact was minimal. This aligns with prior research showing that time pressure alone often falls short of inducing significant stress responses, particularly among experienced individuals, unless combined with high-stakes outcomes or contextual disruptions \cite{zhang2024good}.

Future studies could explore a multi-stressor approach, integrating time constraints with performance evaluations or unpredictable interruptions \cite{perez2025examining}, to better replicate real-world stress conditions.
%Fture studies could adopt a multi-stressor framework, combining time reductions with performance reviews or unpredictable task interruptions \cite{perez2025examining}, to better simulate real-world stress conditions

\begin{summarybox}
    Key takeaway 1: Time pressure alone may be insufficient; combine multiple stressors for more reliable stress induction.
\end{summarybox}

%The lack of consequences in our experimental design likely dimished participants' motivation to experience stress. Unlike real-world programming environments---where performance directly impacts career, professional reputation, or project outcomes, our tasks lacked personal stakes. This aligns with findings that stress responses are amplified when tasks are tied to high-stakes outcomes (e.g., evaluations, rewards, or social judgment) rather than isolated time constraints \cite{}. For instance, studies using the Trier Social Stress Test (TSST) demonstrate that psychosocial stressors like performance evaluation and social task elicit robust physiological and subjective stress responses.\cite{westerink2020deriving}

The variability in individual self-reported stress responses (illustrated in Figure \ref{fig:stress_participant}) highlights the importance of calibrating stress induction protocols through pilot testing. 
Participants who quickly understood tasks and found them engaging reported different experiences than those who struggled. This phenomenon mirrors what Csikszentmihalyi described as the ``flow state,''~\cite{csikszentmihalyi2014flow} where optimal engagement occurs when challenge levels match individual skills. Peifer et al. \cite{peifer2014relation} further established connections between flow experiences and moderate stress levels, suggesting an inverted U-shaped relationship between stress and performance. Our findings suggest that matching the task complexity to participants' skill levels might help ensure enough challenge levels, thus inducing stress.

\begin{summarybox}
    Key takeaway 2: Adapt task difficulty to the skill of the participants.
\end{summarybox}

%Finally, methodological challenges—such as sensor-related data loss—emphasize the importance of pilot studies to validate instrumentation robustness. Pilot testing can uncover flaws in stress induction protocols or task design before full-scale deployment.

%\begin{summarybox}
%Key takeaway: Pilot testing is essential to mitigate technical issues and refine stress induction protocols in complex experiments.
%\end{summarybox}

%\daniela{Something else?}

%What we have learned to inform decisions.

%How to design better stress-inducing protocols.

%Importance of external factors in triggering stress or design considerations for future studies.

%Methodological challenges.

%how perceived relevance, consequence, or engagement affects emotional and physiological responses.

%maybe move this to the next sub-subsection 
%Ethical considerations when introducing stronger stressors 

\subsubsection{\textbf{Ethical Reflections on Stress Induction in Research}}

Inducing stress in controlled experiments presents a fundamental ethical tension: balancing methodological rigour with participant well-being. We employed a protocol to induce stress in participants, enabling its manipulation as an independent variable. Combining biometric (EEG, EDA) and psychometric measurements, we aimed to establish causal relationships between stress and software engineering task performance while reducing reliance on self-reported data. 

Following ethical guidelines to avoid harm and long-term negative effects to our participants \cite{ferreira2019emotional}, we induced moderate, short-term stress (enough to observe effects without harming participants) by limiting the time for the second task. However, there was a risk that this stressor might be insufficient to produce measurable results. This was exactly what happened; our results did not show any stress in our data, meaning that our protocol may have fallen below the threshold needed for observable impact. 

Adding to the previous, another challenge is that individuals perceive and respond to the same stressors differently. This is evident in Figure \ref{fig:stress_participant}, for example, there is no clear pattern in our participants' responses to the stressor. We took this variability to reinforce the importance of post-experiment care and implemented the relaxation video at the end of the experiment to mitigate short-term discomfort \cite{fahey2024importance}.

Furthermore, ethical guidelines also demand careful cost-benefit analysis~\cite{fahey2024importance}. In this case, we might need to create more or harder stressors to get different results; however, imposing discomfort on our participants and offering them little or no immediate benefit will violate the ethical guidelines \cite{apa2017ethics}. Our results denote that ethically constrained stress induction in programming experiments may be fundamentally untenable, either too weak to yield actionable data (as here) or so intense it crosses ethical boundaries.

Finally, our null results invite reflection on the trade-offs between ethical boundaries and experimental validity. Future work could explore alternative stressors (e.g., time pressure in real work environments). 

%if more is needed, to check these websites:
%https://louis.pressbooks.pub/intropsychology/chapter/ethics/

%https://www.organisatiegids.universiteitleiden.nl/binaries/content/assets/sociale-wetenschappen/psychologie/organisatiegids/guidelines-of-the-psychology-research-ethics-committee-april2020.pdf

\subsection{Threats to Validity}

\subsubsection{Internal Validity}
We employed a within-subject design without a control group, which limits the ability to draw causal conclusions. Additionally, there are confounding variables that we could not control and might have affected the overall results. For example, participants' prior experience with programming tasks, fatigue, long-term stress or stress unrelated to the task, could have influenced subjective and biometric responses. We tried to mitigate this using data triangulation, specifically adding an interview at the end to get participants' experiences and impressions.

Furthermore, the participants' motivation to finish or develop the task successfully could have been influenced by the lack of consequences for failure. Motivation is a highly complex variable to control without crossing ethical boundaries. We tried to mitigate this threat by explaining to the participants the importance of completing the tasks in time. 

\subsubsection{External Validity}
This study was thought to be a pilot for future interventions in companies. Hence, we are aware that it is challenging to generalise findings. The number of participants is limited, and they all come from a specific population (PhD students from one university). Future studies need to recruit a more varied sample, including different levels of academic experience, diverse genders and backgrounds, and professionals from the industry to account for a more representative population. Additionally, the artificial nature of lab-like task environments may not fully replicate real-world programming stressors (further discussed in the Ecological validity section).

\subsubsection{Construct Validity}

Our small sample size limits statistical power, especially when interpreting correlations or changes across time points (for example, when comparing the change from Task 1 to Task 2). This issue was compounded by the loss of EEG data points, which reduced the sample size. While the data triangulation adds credibility, conclusions about the efficacy of biometric stress measures or the interpretability of psychometric data must be made cautiously. There is also a risk of confirmation bias in interpreting the alignment between methods, especially when expected outcomes may unconsciously influence how data is coded or analysed.

Moreover, the type of sensors we used in this study impacted the amount and quality of collected data, specifically the EEG data. The sensor was not entirely reliable, and we lost several data points. Additionally, biometric data is influenced by environmental noise, physical movement, or individual physiology \cite{alonso2011quality}.

\subsubsection{Ecological Validity}

Although the study attempted to mimic realistic programming tasks in a daily work scenario, the experimental setting may have induced behaviour not reflective of natural work environments (e.g., being observed or monitored may have influenced stress levels, as one participant commented in the interview). Participants may also have responded differently, knowing there were no consequences if they did not complete the assignment. This lack of real-world accountability may have reduced the urgency or perceived importance of the task, potentially leading to lower stress levels than would be experienced in high-stakes professional contexts. Consequently, the emotional and cognitive responses observed in the study may not fully represent the stress experienced during typical workday demands, deadlines, or performance pressures.

\section{Conclusion}
\label{sec:conclusion}

\subsection{Summary}

In this article, we presented an experimental study to compare psychometric stress measures and biometric indicators and identify stress-related patterns in biometric data during software engineering tasks.

Ten participants wearing biometric sensors performed two tasks, whereby the second task had a stricter time limit.

This limitation did not stress the participants significantly, so our results remain only indicative in terms of confirming or refuting the validity of a comparison of psychometric, biometric, and qualitative data.

\subsection{Future Work}

%\birgit{add ideas: industry replication, another experiment with larger cohort}

We are considering three lines of future work:

\begin{enumerate}
    \item \textbf{Replication:} We are planning to replicate this study with a larger group of software developer participants in industry.
    \item \textbf{Personality and Experiences:} We are curious to explore how individual differences (e.g., personality traits, prior experiences) influence stress response and coping mechanisms.
    \item \textbf{Stress, Motivation \& Performance:} We are designing a study to investigate the relationship between stress, motivation, and performance.
\end{enumerate}

% How do individual differences (e.g., personality traits, prior experiences) influence stress response and coping mechanisms?

% Relationship between stress, motivation, and performance.

\section{Declarations}

\subsection{Funding:} 
The research of Daniela Grassi is partially funded by D.M. 352/2022, Next Generation EU - PNRR, in the scope of the project ``Recognition of emotions of cognitive workers using non-invasive biometric sensors", co-supported by Exprivia, CUP H91I22000410007.
This research was co-funded by the NRRP Initiative, Mission 4, Component 2, Investment 1.3 - Partnerships extended to universities, research centres, companies and research D.D. MUR n. 341, 15.03.2022 – Next Generation EU (``FAIR - Future Artificial Intelligence Research", code PE00000013, CUP H97G22000210007), and the Complementary National Plan PNC-I.1 - Research initiatives for innovative technologies and pathways in the health and welfare sector - D.D. 931 of 06/06/2022 (``DARE - DigitAl lifelong pRevEntion initiative", code PNC0000002, CUP B53C22006420001).

\subsection{Ethical approval:} 
At the time of planning the experiment, formal ethical approval was not required for this type of study. Since then, new procedures have been introduced, and we have submitted our application accordingly. Following a positive preliminary review, we are currently awaiting final approval. Protocol ID: CER\_19720E5F292

\subsection{Informed consent:}
We obtained signed informed consent for all participants in the study.

\subsection{Author Contributions [all authors should be mentioned]}

\textbf{Cristina Martinez Montes:} Conceptualisation, Methodology, Validation, Formal analysis, Investigation, Data curation, Writing - Original Draft, Writing - Review and Editing, Visualization, Project administration.

\textbf{Daniela Grassi:} Conceptualisation, Methodology, Validation, Formal analysis, Investigation, Data curation, Writing - Original Draft, Writing - Review and Editing, Visualization.

\textbf{Nicole Novielli:} Methodology, Validation, Writing - Original Draft, Writing - Review and Editing, Supervision, Project administration, Funding acquisition

\textbf{Birgit Penzenstadler:} Methodology, Validation, Writing - Original Draft, Writing - Review and Editing, Supervision, Project administration, Funding acquisition

\subsection{Data Availability Statement}

The anonymised quantitative data collected are available in our repository:  https://doi.org/10.5281/zenodo.15497559. Qualitative data is not openly available due to reasons of sensitivity.

\subsection{Conflict of Interest}
The authors declare that they have no conflict of interest.

\subsection{Clinical Trial Number in the manuscript.}
Clinical trial number: not applicable.

\section*{Acknowledgements} We thank Robert Feldt for contributions in designing this study and for supportive discussions during the analysis procedure.
We thank our study participants for their time and effort.



\bibliography{bibliography}


\begin{thebibliography}{74}
\ifx \bisbn   \undefined \def \bisbn  #1{ISBN #1}\fi
\ifx \binits  \undefined \def \binits#1{#1}\fi
\ifx \bauthor  \undefined \def \bauthor#1{#1}\fi
\ifx \batitle  \undefined \def \batitle#1{#1}\fi
\ifx \bjtitle  \undefined \def \bjtitle#1{#1}\fi
\ifx \bvolume  \undefined \def \bvolume#1{\textbf{#1}}\fi
\ifx \byear  \undefined \def \byear#1{#1}\fi
\ifx \bissue  \undefined \def \bissue#1{#1}\fi
\ifx \bfpage  \undefined \def \bfpage#1{#1}\fi
\ifx \blpage  \undefined \def \blpage #1{#1}\fi
\ifx \burl  \undefined \def \burl#1{\textsf{#1}}\fi
\ifx \doiurl  \undefined \def \doiurl#1{\url{https://doi.org/#1}}\fi
\ifx \betal  \undefined \def \betal{\textit{et al.}}\fi
\ifx \binstitute  \undefined \def \binstitute#1{#1}\fi
\ifx \binstitutionaled  \undefined \def \binstitutionaled#1{#1}\fi
\ifx \bctitle  \undefined \def \bctitle#1{#1}\fi
\ifx \beditor  \undefined \def \beditor#1{#1}\fi
\ifx \bpublisher  \undefined \def \bpublisher#1{#1}\fi
\ifx \bbtitle  \undefined \def \bbtitle#1{#1}\fi
\ifx \bedition  \undefined \def \bedition#1{#1}\fi
\ifx \bseriesno  \undefined \def \bseriesno#1{#1}\fi
\ifx \blocation  \undefined \def \blocation#1{#1}\fi
\ifx \bsertitle  \undefined \def \bsertitle#1{#1}\fi
\ifx \bsnm \undefined \def \bsnm#1{#1}\fi
\ifx \bsuffix \undefined \def \bsuffix#1{#1}\fi
\ifx \bparticle \undefined \def \bparticle#1{#1}\fi
\ifx \barticle \undefined \def \barticle#1{#1}\fi
\bibcommenthead
\ifx \bconfdate \undefined \def \bconfdate #1{#1}\fi
\ifx \botherref \undefined \def \botherref #1{#1}\fi
\ifx \url \undefined \def \url#1{\textsf{#1}}\fi
\ifx \bchapter \undefined \def \bchapter#1{#1}\fi
\ifx \bbook \undefined \def \bbook#1{#1}\fi
\ifx \bcomment \undefined \def \bcomment#1{#1}\fi
\ifx \oauthor \undefined \def \oauthor#1{#1}\fi
\ifx \citeauthoryear \undefined \def \citeauthoryear#1{#1}\fi
\ifx \endbibitem  \undefined \def \endbibitem {}\fi
\ifx \bconflocation  \undefined \def \bconflocation#1{#1}\fi
\ifx \arxivurl  \undefined \def \arxivurl#1{\textsf{#1}}\fi
\csname PreBibitemsHook\endcsname

%%% 1
\bibitem[\protect\citeauthoryear{Godliauskas and {\v{S}}mite}{2025}]{godliauskas2025well}
\begin{barticle}
\bauthor{\bsnm{Godliauskas}, \binits{P.}},
\bauthor{\bsnm{{\v{S}}mite}, \binits{D.}}:
\batitle{The well-being of software engineers: a systematic literature review and a theory}.
\bjtitle{Empirical Software Engineering}
\bvolume{30}(\bissue{1}),
\bfpage{1}--\blpage{42}
(\byear{2025})
\end{barticle}
\endbibitem

%%% 2
\bibitem[\protect\citeauthoryear{Maslach et~al.}{2001}]{maslach2001job}
\begin{barticle}
\bauthor{\bsnm{Maslach}, \binits{C.}},
\bauthor{\bsnm{Schaufeli}, \binits{W.B.}},
\bauthor{\bsnm{Leiter}, \binits{M.P.}}:
\batitle{Job burnout}.
\bjtitle{Annual review of psychology}
\bvolume{52}(\bissue{2001}),
\bfpage{397}--\blpage{422}
(\byear{2001})
\end{barticle}
\endbibitem

%%% 3
\bibitem[\protect\citeauthoryear{Graziotin et~al.}{2014}]{graziotin2014happy}
\begin{barticle}
\bauthor{\bsnm{Graziotin}, \binits{D.}},
\bauthor{\bsnm{Wang}, \binits{X.}},
\bauthor{\bsnm{Abrahamsson}, \binits{P.}}:
\batitle{Happy software developers solve problems better: psychological measurements in empirical software engineering}.
\bjtitle{PeerJ}
\bvolume{2},
\bfpage{289}
(\byear{2014})
\end{barticle}
\endbibitem

%%% 4
\bibitem[\protect\citeauthoryear{Graziotin and Fagerholm}{2019}]{graziotin2019happiness}
\begin{bchapter}
\bauthor{\bsnm{Graziotin}, \binits{D.}},
\bauthor{\bsnm{Fagerholm}, \binits{F.}}:
\bctitle{Happiness and the productivity of software engineers}.
In: \bbtitle{Rethinking Productivity in Software Engineering},
pp. \bfpage{109}--\blpage{124}.
\bpublisher{Springer}, \blocation{???}
(\byear{2019})
\end{bchapter}
\endbibitem

%%% 5
\bibitem[\protect\citeauthoryear{Bern{\'a}rdez et~al.}{2023}]{bernardez2023empirical}
\begin{barticle}
\bauthor{\bsnm{Bern{\'a}rdez}, \binits{B.}},
\bauthor{\bsnm{Panach}, \binits{J.I.}},
\bauthor{\bsnm{Parejo}, \binits{J.A.}},
\bauthor{\bsnm{Dur{\'a}n}, \binits{A.}},
\bauthor{\bsnm{Juristo}, \binits{N.}},
\bauthor{\bsnm{Ruiz-Cort{\'e}s}, \binits{A.}}:
\batitle{An empirical study to evaluate the impact of mindfulness on helpdesk employees}.
\bjtitle{Science of Computer Programming}
\bvolume{230},
\bfpage{102977}
(\byear{2023})
\end{barticle}
\endbibitem

%%% 6
\bibitem[\protect\citeauthoryear{Montes et~al.}{2024}]{montes2024qualifying}
\begin{bchapter}
\bauthor{\bsnm{Montes}, \binits{C.M.}},
\bauthor{\bsnm{Sj{\"o}gren}, \binits{F.}},
\bauthor{\bsnm{Klevfors}, \binits{A.}},
\bauthor{\bsnm{Penzenstadler}, \binits{B.}}:
\bctitle{Qualifying and quantifying the benefits of mindfulness practices for it workers}.
In: \bbtitle{2024 10th International Conference on ICT for Sustainability (ICT4S)},
pp. \bfpage{272}--\blpage{281}
(\byear{2024}).
\bcomment{IEEE}
\end{bchapter}
\endbibitem

%%% 7
\bibitem[\protect\citeauthoryear{Montes and Penzenstadler}{2023}]{montes2023piloting}
\begin{bchapter}
\bauthor{\bsnm{Montes}, \binits{C.M.}},
\bauthor{\bsnm{Penzenstadler}, \binits{B.}}:
\bctitle{Piloting a well-being and resilience intervention in a course on digitalization for sustainability.}
In: \bbtitle{ICT4S (Doctoral Symposium, Demos, Posters)},
pp. \bfpage{105}--\blpage{118}
(\byear{2023})
\end{bchapter}
\endbibitem

%%% 8
\bibitem[\protect\citeauthoryear{Penzenstadler et~al.}{2022}]{penzenstadler2022take}
\begin{barticle}
\bauthor{\bsnm{Penzenstadler}, \binits{B.}},
\bauthor{\bsnm{Torkar}, \binits{R.}},
\bauthor{\bsnm{Martinez~Montes}, \binits{C.}}:
\batitle{Take a deep breath: Benefits of neuroplasticity practices for software developers and computer workers in a family of experiments}.
\bjtitle{Empirical Software Engineering}
\bvolume{27}(\bissue{4}),
\bfpage{98}
(\byear{2022})
\end{barticle}
\endbibitem

%%% 9
\bibitem[\protect\citeauthoryear{Kreitchmann et~al.}{2019}]{kreitchmann2019controlling}
\begin{barticle}
\bauthor{\bsnm{Kreitchmann}, \binits{R.S.}},
\bauthor{\bsnm{Abad}, \binits{F.J.}},
\bauthor{\bsnm{Ponsoda}, \binits{V.}},
\bauthor{\bsnm{Nieto}, \binits{M.D.}},
\bauthor{\bsnm{Morillo}, \binits{D.}}:
\batitle{Controlling for response biases in self-report scales: Forced-choice vs. psychometric modeling of likert items}.
\bjtitle{Frontiers in psychology}
\bvolume{10},
\bfpage{2309}
(\byear{2019})
\end{barticle}
\endbibitem

%%% 10
\bibitem[\protect\citeauthoryear{Kuncel and Tellegen}{2009}]{kuncel2009conceptual}
\begin{barticle}
\bauthor{\bsnm{Kuncel}, \binits{N.R.}},
\bauthor{\bsnm{Tellegen}, \binits{A.}}:
\batitle{A conceptual and empirical reexamination of the measurement of the social desirability of items: Implications for detecting desirable response style and scale development}.
\bjtitle{Personnel Psychology}
\bvolume{62}(\bissue{2}),
\bfpage{201}--\blpage{228}
(\byear{2009})
\end{barticle}
\endbibitem

%%% 11
\bibitem[\protect\citeauthoryear{Weijters et~al.}{2013}]{weijters2013reversed}
\begin{barticle}
\bauthor{\bsnm{Weijters}, \binits{B.}},
\bauthor{\bsnm{Baumgartner}, \binits{H.}},
\bauthor{\bsnm{Schillewaert}, \binits{N.}}:
\batitle{Reversed item bias: an integrative model.}
\bjtitle{Psychological methods}
\bvolume{18}(\bissue{3}),
\bfpage{320}
(\byear{2013})
\end{barticle}
\endbibitem

%%% 12
\bibitem[\protect\citeauthoryear{{M{\"u}ller} and {Fritz}}{2015}]{MF15}
\begin{bchapter}
\bauthor{\bsnm{{M{\"u}ller}}, \binits{S.C.}},
\bauthor{\bsnm{{Fritz}}, \binits{T.}}:
\bctitle{Stuck and frustrated or in flow and happy: Sensing developers' emotions and progress}.
In: \bbtitle{ICSE},
pp. \bfpage{688}--\blpage{699}
(\byear{2015})
\end{bchapter}
\endbibitem

%%% 13
\bibitem[\protect\citeauthoryear{Vrzakova et~al.}{2020}]{Vrzakova2020}
\begin{botherref}
\oauthor{\bsnm{Vrzakova}, \binits{H.}},
\oauthor{\bsnm{Begel}, \binits{A.}},
\oauthor{\bsnm{Mehtätalo}, \binits{L.}},
\oauthor{\bsnm{Bednarik}, \binits{R.}}:
Affect recognition in code review: An in-situ biometric study of reviewer’s affect.
J. Syst. Softw.
\textbf{159}
(2020)
\doiurl{10.1016/j.jss.2019.110434}
\end{botherref}
\endbibitem

%%% 14
\bibitem[\protect\citeauthoryear{Girardi et~al.}{2020}]{girardi2020recognizing}
\begin{bchapter}
\bauthor{\bsnm{Girardi}, \binits{D.}},
\bauthor{\bsnm{Novielli}, \binits{N.}},
\bauthor{\bsnm{Fucci}, \binits{D.}},
\bauthor{\bsnm{Lanubile}, \binits{F.}}:
\bctitle{Recognizing developers' emotions while programming}.
In: \bbtitle{Proceedings of the ACM/IEEE 42nd International Conference on Software Engineering},
pp. \bfpage{666}--\blpage{677}
(\byear{2020})
\end{bchapter}
\endbibitem

%%% 15
\bibitem[\protect\citeauthoryear{Grassi et~al.}{2025}]{Grassi_etAl:2025}
\begin{bchapter}
\bauthor{\bsnm{Grassi}, \binits{D.}},
\bauthor{\bsnm{Lanubile}, \binits{F.}},
\bauthor{\bsnm{Motca-Schnabel}, \binits{A.}},
\bauthor{\bsnm{Novielli}, \binits{N.}}:
\bctitle{A cluster-based approach for emotion recognition in software development}.
In: \bbtitle{Proceedings of the 18th International Conference on Cooperative and Human Aspects of Software Engineering (CHASE 2025)},
pp. \bfpage{1}--\blpage{13}
(\byear{2025}).
\doiurl{10.1109/CHASE66643.2025.00034}
\end{bchapter}
\endbibitem

%%% 16
\bibitem[\protect\citeauthoryear{Russell}{1980}]{russell1980circumplex}
\begin{barticle}
\bauthor{\bsnm{Russell}, \binits{J.A.}}:
\batitle{A circumplex model of affect.}
\bjtitle{Journal of personality and social psychology}
\bvolume{39}(\bissue{6}),
\bfpage{1161}
(\byear{1980})
\end{barticle}
\endbibitem

%%% 17
\bibitem[\protect\citeauthoryear{Westerink et~al.}{2020}]{westerink2020deriving}
\begin{barticle}
\bauthor{\bsnm{Westerink}, \binits{J.H.}},
\bauthor{\bsnm{Rajae-Joordens}, \binits{R.J.}},
\bauthor{\bsnm{Ouwerkerk}, \binits{M.}},
\bauthor{\bsnm{Dooren}, \binits{M.}},
\bauthor{\bsnm{Jelfs}, \binits{S.}},
\bauthor{\bsnm{Denissen}, \binits{A.J.}},
\bauthor{\bsnm{Vries}, \binits{E.}},
\bauthor{\bsnm{Ee}, \binits{R.}}:
\batitle{Deriving a cortisol-related stress indicator from wearable skin conductance measurements: Quantitative model \& experimental validation}.
\bjtitle{Frontiers in Computer Science}
\bvolume{2},
\bfpage{39}
(\byear{2020})
\end{barticle}
\endbibitem

%%% 18
\bibitem[\protect\citeauthoryear{Kocielnik et~al.}{2013}]{Kocielnik}
\begin{bchapter}
\bauthor{\bsnm{Kocielnik}, \binits{R.}},
\bauthor{\bsnm{Sidorova}, \binits{N.}},
\bauthor{\bsnm{Maggi}, \binits{F.M.}},
\bauthor{\bsnm{Ouwerkerk}, \binits{M.}},
\bauthor{\bsnm{Westerink}, \binits{J.H.D.M.}}:
\bctitle{Smart technologies for long-term stress monitoring at work}.
In: \bbtitle{Proceedings of the 26th IEEE International Symposium on Computer-Based Medical Systems},
pp. \bfpage{53}--\blpage{58}
(\byear{2013}).
\doiurl{10.1109/CBMS.2013.6627764}
\end{bchapter}
\endbibitem

%%% 19
\bibitem[\protect\citeauthoryear{Fan et~al.}{2015}]{fan:stress:anxiety}
\begin{barticle}
\bauthor{\bsnm{Fan}, \binits{L.-B.}},
\bauthor{\bsnm{Blumenthal}, \binits{J.A.}},
\bauthor{\bsnm{Watkins}, \binits{L.L.}},
\bauthor{\bsnm{Sherwood}, \binits{A.}}:
\batitle{{Work and home stress: associations with anxiety and depression symptoms}}.
\bjtitle{Occupational Medicine}
\bvolume{65}(\bissue{2}),
\bfpage{110}--\blpage{116}
(\byear{2015})
\doiurl{10.1093/occmed/kqu181}
{\href{https://arxiv.org/abs/https://academic.oup.com/occmed/article-pdf/65/2/110/4312120/kqu181.pdf}{{https://academic.oup.com/occmed/article-pdf/65/2/110/4312120/kqu181.pdf}}}
\end{barticle}
\endbibitem

%%% 20
\bibitem[\protect\citeauthoryear{Posner et~al.}{2005}]{Russell2005}
\begin{barticle}
\bauthor{\bsnm{Posner}, \binits{J.}},
\bauthor{\bsnm{Russell}, \binits{J.}},
\bauthor{\bsnm{Peterson}, \binits{B.}}:
\batitle{The circumplex model of affect: an integrative approach to affective neuroscience, cognitive development, and psychopathology.}
\bjtitle{Dev Psychopathol.}
\bvolume{17(3)},
\bfpage{715}--\blpage{34}
(\byear{2005})
\doiurl{10.1017/S095457940505034}
\end{barticle}
\endbibitem

%%% 21
\bibitem[\protect\citeauthoryear{{American Psychological Association}}{2018}]{apa2018dictionary}
\begin{bbook}
\bauthor{\bsnm{{American Psychological Association}}}:
\bbtitle{APA Dictionary of Psychology}.
\bpublisher{American Psychological Association},
\blocation{Washington, DC}
(\byear{2018}).
\burl{https://dictionary.apa.org}
\end{bbook}
\endbibitem

%%% 22
\bibitem[\protect\citeauthoryear{Kim and Andr{\'{e}}}{2008}]{KimA08}
\begin{barticle}
\bauthor{\bsnm{Kim}, \binits{J.}},
\bauthor{\bsnm{Andr{\'{e}}}, \binits{E.}}:
\batitle{Emotion recognition based on physiological changes in music listening}.
\bjtitle{{IEEE} Trans. on Pattern Analysis and Machine Intelligence}
\bvolume{30}(\bissue{12}),
\bfpage{2067}--\blpage{2083}
(\byear{2008})
\doiurl{10.1109/TPAMI.2008.26}
\end{barticle}
\endbibitem

%%% 23
\bibitem[\protect\citeauthoryear{Kim et~al.}{2004}]{kim2004emotion}
\begin{barticle}
\bauthor{\bsnm{Kim}, \binits{K.H.}},
\bauthor{\bsnm{Bang}, \binits{S.W.}},
\bauthor{\bsnm{Kim}, \binits{S.R.}}:
\batitle{Emotion recognition system using short-term monitoring of physiological signals}.
\bjtitle{Medical and biological engineering and computing}
\bvolume{42},
\bfpage{419}--\blpage{427}
(\byear{2004})
\end{barticle}
\endbibitem

%%% 24
\bibitem[\protect\citeauthoryear{Soleymani et~al.}{2015}]{soleymani2015analysis}
\begin{barticle}
\bauthor{\bsnm{Soleymani}, \binits{M.}},
\bauthor{\bsnm{Asghari-Esfeden}, \binits{S.}},
\bauthor{\bsnm{Fu}, \binits{Y.}},
\bauthor{\bsnm{Pantic}, \binits{M.}}:
\batitle{Analysis of eeg signals and facial expressions for continuous emotion detection}.
\bjtitle{IEEE Transactions on Affective Computing}
\bvolume{7}(\bissue{1}),
\bfpage{17}--\blpage{28}
(\byear{2015})
\end{barticle}
\endbibitem

%%% 25
\bibitem[\protect\citeauthoryear{Koelstra et~al.}{2012}]{KoelstraMSLYEPNP12}
\begin{barticle}
\bauthor{\bsnm{Koelstra}, \binits{S.}},
\bauthor{\bsnm{M{\"{u}}hl}, \binits{C.}},
\bauthor{\bsnm{Soleymani}, \binits{M.}},
\bauthor{\bsnm{Lee}, \binits{J.}},
\bauthor{\bsnm{Yazdani}, \binits{A.}},
\bauthor{\bsnm{Ebrahimi}, \binits{T.}},
\bauthor{\bsnm{Pun}, \binits{T.}},
\bauthor{\bsnm{Nijholt}, \binits{A.}},
\bauthor{\bsnm{Patras}, \binits{I.}}:
\batitle{{DEAP:} {A} database for emotion analysis using physiological signals}.
\bjtitle{{IEEE} Trans. on Affective Comp.}
\bvolume{3}(\bissue{1}),
\bfpage{18}--\blpage{31}
(\byear{2012})
\doiurl{10.1109/T-AFFC.2011.15}
\end{barticle}
\endbibitem

%%% 26
\bibitem[\protect\citeauthoryear{Khan et~al.}{2011}]{Khan_debug_emotion}
\begin{barticle}
\bauthor{\bsnm{Khan}, \binits{I.A.}},
\bauthor{\bsnm{Brinkman}, \binits{W.-P.}},
\bauthor{\bsnm{Hierons}, \binits{R.M.}}:
\batitle{Do moods affect programmers’ debug performance?}
\bjtitle{Cogn. Technol. Work}
\bvolume{13}(\bissue{4}),
\bfpage{245}--\blpage{258}
(\byear{2011})
\doiurl{10.1007/s10111-010-0164-1}
\end{barticle}
\endbibitem

%%% 27
\bibitem[\protect\citeauthoryear{Russell}{1980}]{Russell1980}
\begin{barticle}
\bauthor{\bsnm{Russell}, \binits{J.}}:
\batitle{A circumplex model of affect}.
\bjtitle{Journal of Personality and Social Psychology}
\bvolume{39},
\bfpage{1161}--\blpage{1178}
(\byear{1980})
\doiurl{10.1037/h0077714}
\end{barticle}
\endbibitem

%%% 28
\bibitem[\protect\citeauthoryear{Goodman et~al.}{2013}]{goodman2013stress}
\begin{barticle}
\bauthor{\bsnm{Goodman}, \binits{R.N.}},
\bauthor{\bsnm{Rietschel}, \binits{J.C.}},
\bauthor{\bsnm{Lo}, \binits{L.-C.}},
\bauthor{\bsnm{Costanzo}, \binits{M.E.}},
\bauthor{\bsnm{Hatfield}, \binits{B.D.}}:
\batitle{Stress, emotion regulation and cognitive performance: The predictive contributions of trait and state relative frontal eeg alpha asymmetry}.
\bjtitle{International journal of psychophysiology}
\bvolume{87}(\bissue{2}),
\bfpage{115}--\blpage{123}
(\byear{2013})
\end{barticle}
\endbibitem

%%% 29
\bibitem[\protect\citeauthoryear{Bradley and Lang}{2000}]{bradley2000measuring}
\begin{botherref}
\oauthor{\bsnm{Bradley}, \binits{M.M.}},
\oauthor{\bsnm{Lang}, \binits{P.J.}}:
Measuring emotion: Behavior, feeling, and physiology.
(2000)
\end{botherref}
\endbibitem

%%% 30
\bibitem[\protect\citeauthoryear{Girardi et~al.}{2021}]{TSE_Girardi}
\begin{barticle}
\bauthor{\bsnm{Girardi}, \binits{D.}},
\bauthor{\bsnm{Lanubile}, \binits{F.}},
\bauthor{\bsnm{Novielli}, \binits{N.}},
\bauthor{\bsnm{Serebrenik}, \binits{A.}}:
\batitle{Emotions and perceived productivity of software developers at the workplace}.
\bjtitle{IEEE Transactions on Software Engineering}
\bvolume{48}(\bissue{9}),
\bfpage{3326}--\blpage{3341}
(\byear{2021})
\end{barticle}
\endbibitem

%%% 31
\bibitem[\protect\citeauthoryear{Canento et~al.}{2011}]{Canento_2011}
\begin{bchapter}
\bauthor{\bsnm{Canento}, \binits{F.}},
\bauthor{\bsnm{Fred}, \binits{A.}},
\bauthor{\bsnm{Silva}, \binits{H.}},
\bauthor{\bsnm{Gamboa}, \binits{H.}},
\bauthor{\bsnm{Lourenço}, \binits{A.}}:
\bctitle{Multimodal biosignal sensor data handling for emotion recognition}.
In: \bbtitle{SENSORS, 2011 IEEE},
pp. \bfpage{647}--\blpage{650}
(\byear{2011}).
\doiurl{10.1109/ICSENS.2011.6127029}
\end{bchapter}
\endbibitem

%%% 32
\bibitem[\protect\citeauthoryear{Castaldo et~al.}{2015}]{castaldo2015acute}
\begin{barticle}
\bauthor{\bsnm{Castaldo}, \binits{R.}},
\bauthor{\bsnm{Melillo}, \binits{P.}},
\bauthor{\bsnm{Bracale}, \binits{U.}},
\bauthor{\bsnm{Caserta}, \binits{M.}},
\bauthor{\bsnm{Triassi}, \binits{M.}},
\bauthor{\bsnm{Pecchia}, \binits{L.}}:
\batitle{Acute mental stress assessment via short term hrv analysis in healthy adults: A systematic review with meta-analysis}.
\bjtitle{Biomedical Signal Processing and Control}
\bvolume{18},
\bfpage{370}--\blpage{377}
(\byear{2015})
\end{barticle}
\endbibitem

%%% 33
\bibitem[\protect\citeauthoryear{Saeed et~al.}{2020}]{saeed2020eeg}
\begin{barticle}
\bauthor{\bsnm{Saeed}, \binits{S.M.U.}},
\bauthor{\bsnm{Anwar}, \binits{S.M.}},
\bauthor{\bsnm{Khalid}, \binits{H.}},
\bauthor{\bsnm{Majid}, \binits{M.}},
\bauthor{\bsnm{Bagci}, \binits{U.}}:
\batitle{Eeg based classification of long-term stress using psychological labeling}.
\bjtitle{Sensors}
\bvolume{20}(\bissue{7}),
\bfpage{1886}
(\byear{2020})
\end{barticle}
\endbibitem

%%% 34
\bibitem[\protect\citeauthoryear{Chae et~al.}{2021}]{chae2021relationship}
\begin{barticle}
\bauthor{\bsnm{Chae}, \binits{J.}},
\bauthor{\bsnm{Hwang}, \binits{S.}},
\bauthor{\bsnm{Seo}, \binits{W.}},
\bauthor{\bsnm{Kang}, \binits{Y.}}:
\batitle{Relationship between rework of engineering drawing tasks and stress level measured from physiological signals}.
\bjtitle{Automation in Construction}
\bvolume{124},
\bfpage{103560}
(\byear{2021})
\end{barticle}
\endbibitem

%%% 35
\bibitem[\protect\citeauthoryear{Mohanavelu et~al.}{2020}]{mohanavelu2020dynamic}
\begin{barticle}
\bauthor{\bsnm{Mohanavelu}, \binits{K.}},
\bauthor{\bsnm{Poonguzhali}, \binits{S.}},
\bauthor{\bsnm{Adalarasu}, \binits{K.}},
\bauthor{\bsnm{Ravi}, \binits{D.}},
\bauthor{\bsnm{Chinnadurai}, \binits{V.}},
\bauthor{\bsnm{Vinutha}, \binits{S.}},
\bauthor{\bsnm{Ramachandran}, \binits{K.}},
\bauthor{\bsnm{Jayaraman}, \binits{S.}}:
\batitle{Dynamic cognitive workload assessment for fighter pilots in simulated fighter aircraft environment using eeg}.
\bjtitle{Biomedical Signal Processing and Control}
\bvolume{61},
\bfpage{102018}
(\byear{2020})
\end{barticle}
\endbibitem

%%% 36
\bibitem[\protect\citeauthoryear{Mart{\'\i}nez~V{\'a}squez et~al.}{2023}]{martinez2023mutual}
\begin{barticle}
\bauthor{\bsnm{Mart{\'\i}nez~V{\'a}squez}, \binits{D.A.}},
\bauthor{\bsnm{Posada-Quintero}, \binits{H.F.}},
\bauthor{\bsnm{Rivera~Pinz{\'o}n}, \binits{D.M.}}:
\batitle{Mutual information between eda and eeg in multiple cognitive tasks and sleep deprivation conditions}.
\bjtitle{Behavioral Sciences}
\bvolume{13}(\bissue{9}),
\bfpage{707}
(\byear{2023})
\end{barticle}
\endbibitem

%%% 37
\bibitem[\protect\citeauthoryear{Fucci et~al.}{2019a}]{Fucci_etAL:ICPC2019}
\begin{bchapter}
\bauthor{\bsnm{Fucci}, \binits{D.}},
\bauthor{\bsnm{Girardi}, \binits{D.}},
\bauthor{\bsnm{Novielli}, \binits{N.}},
\bauthor{\bsnm{Quaranta}, \binits{L.}},
\bauthor{\bsnm{Lanubile}, \binits{F.}}:
\bctitle{A replication study on code comprehension and expertise using lightweight biometric sensors}.
In: \bbtitle{2019 IEEE/ACM 27th International Conference on Program Comprehension (ICPC)},
pp. \bfpage{311}--\blpage{322}
(\byear{2019}).
\doiurl{10.1109/ICPC.2019.00050}
\end{bchapter}
\endbibitem

%%% 38
\bibitem[\protect\citeauthoryear{Fucci et~al.}{2019b}]{fucci2019replication}
\begin{bchapter}
\bauthor{\bsnm{Fucci}, \binits{D.}},
\bauthor{\bsnm{Girardi}, \binits{D.}},
\bauthor{\bsnm{Novielli}, \binits{N.}},
\bauthor{\bsnm{Quaranta}, \binits{L.}},
\bauthor{\bsnm{Lanubile}, \binits{F.}}:
\bctitle{A replication study on code comprehension and expertise using lightweight biometric sensors}.
In: \bbtitle{2019 IEEE/ACM 27th International Conference on Program Comprehension (ICPC)},
pp. \bfpage{311}--\blpage{322}
(\byear{2019}).
\bcomment{IEEE}
\end{bchapter}
\endbibitem

%%% 39
\bibitem[\protect\citeauthoryear{Radevski et~al.}{2015}]{RHM15}
\begin{bchapter}
\bauthor{\bsnm{Radevski}, \binits{S.}},
\bauthor{\bsnm{Hata}, \binits{H.}},
\bauthor{\bsnm{Matsumoto}, \binits{K.}}:
\bctitle{{Real-time Monitoring of Neural State in Assessing and Improving Software Developers' Productivity}}.
In: \bbtitle{Proceedings of the Eighth International Workshop on Cooperative and Human Aspects of Software Engineering},
pp. \bfpage{93}--\blpage{96}
(\byear{2015}).
\bcomment{IEEE Press}
\end{bchapter}
\endbibitem

%%% 40
\bibitem[\protect\citeauthoryear{Z{\"u}ger et~al.}{2017}]{ZCM17}
\begin{bchapter}
\bauthor{\bsnm{Z{\"u}ger}, \binits{M.}},
\bauthor{\bsnm{Corley}, \binits{C.}},
\bauthor{\bsnm{Meyer}, \binits{A.N.}},
\bauthor{\bsnm{Li}, \binits{B.}},
\bauthor{\bsnm{Fritz}, \binits{T.}},
\bauthor{\bsnm{Shepherd}, \binits{D.}},
\bauthor{\bsnm{Augustine}, \binits{V.}},
\bauthor{\bsnm{Francis}, \binits{P.}},
\bauthor{\bsnm{Kraft}, \binits{N.}},
\bauthor{\bsnm{Snipes}, \binits{W.}}:
\bctitle{{Reducing Interruptions at Work: A Large-scale Field Study of FlowLight}}.
In: \bbtitle{Proceedings of the 2017 CHI Conference on Human Factors in Computing Systems},
pp. \bfpage{61}--\blpage{72}
(\byear{2017}).
\bcomment{ACM}
\end{bchapter}
\endbibitem

%%% 41
\bibitem[\protect\citeauthoryear{Calcagno et~al.}{2020}]{calcagno2020eeg}
\begin{bchapter}
\bauthor{\bsnm{Calcagno}, \binits{A.}},
\bauthor{\bsnm{Coelli}, \binits{S.}},
\bauthor{\bsnm{Couceiro}, \binits{R.}},
\bauthor{\bsnm{Dur{\~a}es}, \binits{J.}},
\bauthor{\bsnm{Amendola}, \binits{C.}},
\bauthor{\bsnm{Pirovano}, \binits{I.}},
\bauthor{\bsnm{Re}, \binits{R.}},
\bauthor{\bsnm{Bianchi}, \binits{A.M.}}:
\bctitle{Eeg monitoring during software development}.
In: \bbtitle{2020 IEEE 20th Mediterranean Electrotechnical Conference (MELECON)},
pp. \bfpage{325}--\blpage{329}
(\byear{2020}).
\bcomment{IEEE}
\end{bchapter}
\endbibitem

%%% 42
\bibitem[\protect\citeauthoryear{Medeiros et~al.}{2021}]{medeiros2021can}
\begin{barticle}
\bauthor{\bsnm{Medeiros}, \binits{J.}},
\bauthor{\bsnm{Couceiro}, \binits{R.}},
\bauthor{\bsnm{Duarte}, \binits{G.}},
\bauthor{\bsnm{Dur{\~a}es}, \binits{J.}},
\bauthor{\bsnm{Castelhano}, \binits{J.}},
\bauthor{\bsnm{Duarte}, \binits{C.}},
\bauthor{\bsnm{Castelo-Branco}, \binits{M.}},
\bauthor{\bsnm{Madeira}, \binits{H.}},
\bauthor{\bsnm{De~Carvalho}, \binits{P.}},
\bauthor{\bsnm{Teixeira}, \binits{C.}}:
\batitle{Can eeg be adopted as a neuroscience reference for assessing software programmers’ cognitive load?}
\bjtitle{Sensors}
\bvolume{21}(\bissue{7}),
\bfpage{2338}
(\byear{2021})
\end{barticle}
\endbibitem

%%% 43
\bibitem[\protect\citeauthoryear{Coan and Allen}{2007}]{coan2007handbook}
\begin{bbook}
\bauthor{\bsnm{Coan}, \binits{J.A.}},
\bauthor{\bsnm{Allen}, \binits{J.J.}}:
\bbtitle{Handbook of Emotion Elicitation and Assessment}.
\bpublisher{Oxford university press}, \blocation{???}
(\byear{2007})
\end{bbook}
\endbibitem

%%% 44
\bibitem[\protect\citeauthoryear{Martinez~Montes et~al.}{2025}]{martinez_montes_cristina_2025_15497559}
\begin{botherref}
\oauthor{\bsnm{Martinez~Montes}, \binits{C.}},
\oauthor{\bsnm{Grassi}, \binits{D.}},
\oauthor{\bsnm{Novielli}, \binits{N.}},
\oauthor{\bsnm{Penzenstadler}, \binits{B.}}:
Replication Package for Stress Multimodal Study.
\url{https://doi.org/10.5281/zenodo.15497559}
\end{botherref}
\endbibitem

%%% 45
\bibitem[\protect\citeauthoryear{Lee}{2012}]{lee2012review}
\begin{barticle}
\bauthor{\bsnm{Lee}, \binits{E.-H.}}:
\batitle{Review of the psychometric evidence of the perceived stress scale}.
\bjtitle{Asian nursing research}
\bvolume{6}(\bissue{4}),
\bfpage{121}--\blpage{127}
(\byear{2012})
\end{barticle}
\endbibitem

%%% 46
\bibitem[\protect\citeauthoryear{Helton and N{\"a}swall}{2015}]{helton2015short}
\begin{botherref}
\oauthor{\bsnm{Helton}, \binits{W.S.}},
\oauthor{\bsnm{N{\"a}swall}, \binits{K.}}:
Short stress state questionnaire.
European Journal of Psychological Assessment
(2015)
\end{botherref}
\endbibitem

%%% 47
\bibitem[\protect\citeauthoryear{Hart}{2006}]{hart2006nasa}
\begin{bchapter}
\bauthor{\bsnm{Hart}, \binits{S.G.}}:
\bctitle{Nasa-task load index (nasa-tlx); 20 years later}.
In: \bbtitle{Proceedings of the Human Factors and Ergonomics Society Annual Meeting},
vol. \bseriesno{50},
pp. \bfpage{904}--\blpage{908}
(\byear{2006}).
\bcomment{Sage publications Sage CA: Los Angeles, CA}
\end{bchapter}
\endbibitem

%%% 48
\bibitem[\protect\citeauthoryear{Taylor et~al.}{2015}]{taylor2015automatic}
\begin{bchapter}
\bauthor{\bsnm{Taylor}, \binits{S.}},
\bauthor{\bsnm{Jaques}, \binits{N.}},
\bauthor{\bsnm{Chen}, \binits{W.}},
\bauthor{\bsnm{Fedor}, \binits{S.}},
\bauthor{\bsnm{Sano}, \binits{A.}},
\bauthor{\bsnm{Picard}, \binits{R.}}:
\bctitle{Automatic identification of artifacts in electrodermal activity data}.
In: \bbtitle{2015 37th Annual International Conference of the IEEE Engineering in Medicine and Biology Society (EMBC)},
pp. \bfpage{1934}--\blpage{1937}
(\byear{2015}).
\bcomment{IEEE}
\end{bchapter}
\endbibitem

%%% 49
\bibitem[\protect\citeauthoryear{Greco et~al.}{2015}]{greco2015cvxeda}
\begin{barticle}
\bauthor{\bsnm{Greco}, \binits{A.}},
\bauthor{\bsnm{Valenza}, \binits{G.}},
\bauthor{\bsnm{Lanata}, \binits{A.}},
\bauthor{\bsnm{Scilingo}, \binits{E.P.}},
\bauthor{\bsnm{Citi}, \binits{L.}}:
\batitle{cvxeda: A convex optimization approach to electrodermal activity processing}.
\bjtitle{IEEE transactions on biomedical engineering}
\bvolume{63}(\bissue{4}),
\bfpage{797}--\blpage{804}
(\byear{2015})
\end{barticle}
\endbibitem

%%% 50
\bibitem[\protect\citeauthoryear{Nardelli et~al.}{2022}]{nardelli2022comeda}
\begin{barticle}
\bauthor{\bsnm{Nardelli}, \binits{M.}},
\bauthor{\bsnm{Greco}, \binits{A.}},
\bauthor{\bsnm{Sebastiani}, \binits{L.}},
\bauthor{\bsnm{Scilingo}, \binits{E.P.}}:
\batitle{Comeda: A new tool for stress assessment based on electrodermal activity}.
\bjtitle{Computers in Biology and Medicine}
\bvolume{150},
\bfpage{106144}
(\byear{2022})
\end{barticle}
\endbibitem

%%% 51
\bibitem[\protect\citeauthoryear{Healey and Picard}{2005}]{stress_picard}
\begin{barticle}
\bauthor{\bsnm{Healey}, \binits{J.A.}},
\bauthor{\bsnm{Picard}, \binits{R.W.}}:
\batitle{Detecting stress during real-world driving tasks using physiological sensors}.
\bjtitle{IEEE Transactions on Intelligent Transportation Systems}
\bvolume{6}(\bissue{2}),
\bfpage{156}--\blpage{166}
(\byear{2005})
\doiurl{10.1109/TITS.2005.848368}
\end{barticle}
\endbibitem

%%% 52
\bibitem[\protect\citeauthoryear{Gulati et~al.}{2010}]{gulati2010heart}
\begin{barticle}
\bauthor{\bsnm{Gulati}, \binits{M.}},
\bauthor{\bsnm{Shaw}, \binits{L.J.}},
\bauthor{\bsnm{Thisted}, \binits{R.A.}},
\bauthor{\bsnm{Black}, \binits{H.R.}},
\bauthor{\bsnm{Bairey~Merz}, \binits{C.N.}},
\bauthor{\bsnm{Arnsdorf}, \binits{M.F.}}:
\batitle{Heart rate response to exercise stress testing in asymptomatic women: the st. james women take heart project}.
\bjtitle{Circulation}
\bvolume{122}(\bissue{2}),
\bfpage{130}--\blpage{137}
(\byear{2010})
\end{barticle}
\endbibitem

%%% 53
\bibitem[\protect\citeauthoryear{Malik and Camm}{1990}]{malik1990heart}
\begin{barticle}
\bauthor{\bsnm{Malik}, \binits{M.}},
\bauthor{\bsnm{Camm}, \binits{A.J.}}:
\batitle{Heart rate variability}.
\bjtitle{Clinical cardiology}
\bvolume{13}(\bissue{8}),
\bfpage{570}--\blpage{576}
(\byear{1990})
\end{barticle}
\endbibitem

%%% 54
\bibitem[\protect\citeauthoryear{Yu et~al.}{2024}]{yu2024exploring}
\begin{barticle}
\bauthor{\bsnm{Yu}, \binits{X.}},
\bauthor{\bsnm{Lu}, \binits{J.}},
\bauthor{\bsnm{Liu}, \binits{W.}},
\bauthor{\bsnm{Cheng}, \binits{Z.}},
\bauthor{\bsnm{Xiao}, \binits{G.}}:
\batitle{Exploring physiological stress response evoked by passive translational acceleration in healthy adults: a pilot study utilizing electrodermal activity and heart rate variability measurements}.
\bjtitle{Scientific Reports}
\bvolume{14}(\bissue{1}),
\bfpage{11349}
(\byear{2024})
\end{barticle}
\endbibitem

%%% 55
\bibitem[\protect\citeauthoryear{Reinhardt et~al.}{2012}]{reinhardt2012salivary}
\begin{barticle}
\bauthor{\bsnm{Reinhardt}, \binits{T.}},
\bauthor{\bsnm{Schmahl}, \binits{C.}},
\bauthor{\bsnm{W{\"u}st}, \binits{S.}},
\bauthor{\bsnm{Bohus}, \binits{M.}}:
\batitle{Salivary cortisol, heart rate, electrodermal activity and subjective stress responses to the mannheim multicomponent stress test (mmst)}.
\bjtitle{Psychiatry research}
\bvolume{198}(\bissue{1}),
\bfpage{106}--\blpage{111}
(\byear{2012})
\end{barticle}
\endbibitem

%%% 56
\bibitem[\protect\citeauthoryear{Delliaux et~al.}{2019}]{delliaux2019mental}
\begin{barticle}
\bauthor{\bsnm{Delliaux}, \binits{S.}},
\bauthor{\bsnm{Delaforge}, \binits{A.}},
\bauthor{\bsnm{Deharo}, \binits{J.-C.}},
\bauthor{\bsnm{Chaumet}, \binits{G.}}:
\batitle{Mental workload alters heart rate variability, lowering non-linear dynamics}.
\bjtitle{Frontiers in physiology}
\bvolume{10},
\bfpage{565}
(\byear{2019})
\end{barticle}
\endbibitem

%%% 57
\bibitem[\protect\citeauthoryear{Braun and Clarke}{2021}]{braun2021thematic}
\begin{botherref}
\oauthor{\bsnm{Braun}, \binits{V.}},
\oauthor{\bsnm{Clarke}, \binits{V.}}:
Thematic analysis: A practical guide
(2021)
\end{botherref}
\endbibitem

%%% 58
\bibitem[\protect\citeauthoryear{Adamson et~al.}{2020}]{adamson2020international}
\begin{barticle}
\bauthor{\bsnm{Adamson}, \binits{M.M.}},
\bauthor{\bsnm{Phillips}, \binits{A.}},
\bauthor{\bsnm{Seenivasan}, \binits{S.}},
\bauthor{\bsnm{Martinez}, \binits{J.}},
\bauthor{\bsnm{Grewal}, \binits{H.}},
\bauthor{\bsnm{Kang}, \binits{X.}},
\bauthor{\bsnm{Coetzee}, \binits{J.}},
\bauthor{\bsnm{Luttenbacher}, \binits{I.}},
\bauthor{\bsnm{Jester}, \binits{A.}},
\bauthor{\bsnm{Harris}, \binits{O.A.}}, \betal:
\batitle{International prevalence and correlates of psychological stress during the global covid-19 pandemic}.
\bjtitle{International Journal of Environmental Research and Public Health}
\bvolume{17}(\bissue{24}),
\bfpage{9248}
(\byear{2020})
\end{barticle}
\endbibitem

%%% 59
\bibitem[\protect\citeauthoryear{Hellhammer et~al.}{2009}]{hellhammer2009salivary}
\begin{barticle}
\bauthor{\bsnm{Hellhammer}, \binits{D.H.}},
\bauthor{\bsnm{W{\"u}st}, \binits{S.}},
\bauthor{\bsnm{Kudielka}, \binits{B.M.}}:
\batitle{Salivary cortisol as a biomarker in stress research}.
\bjtitle{Psychoneuroendocrinology}
\bvolume{34}(\bissue{2}),
\bfpage{163}--\blpage{171}
(\byear{2009})
\end{barticle}
\endbibitem

%%% 60
\bibitem[\protect\citeauthoryear{Thayer et~al.}{2012}]{thayer2012meta}
\begin{barticle}
\bauthor{\bsnm{Thayer}, \binits{J.F.}},
\bauthor{\bsnm{{\AA}hs}, \binits{F.}},
\bauthor{\bsnm{Fredrikson}, \binits{M.}},
\bauthor{\bsnm{Sollers~III}, \binits{J.J.}},
\bauthor{\bsnm{Wager}, \binits{T.D.}}:
\batitle{A meta-analysis of heart rate variability and neuroimaging studies: implications for heart rate variability as a marker of stress and health}.
\bjtitle{Neuroscience \& Biobehavioral Reviews}
\bvolume{36}(\bissue{2}),
\bfpage{747}--\blpage{756}
(\byear{2012})
\end{barticle}
\endbibitem

%%% 61
\bibitem[\protect\citeauthoryear{Sandner et~al.}{2020}]{sandner2020investigating}
\begin{barticle}
\bauthor{\bsnm{Sandner}, \binits{M.}},
\bauthor{\bsnm{Lois}, \binits{G.}},
\bauthor{\bsnm{Streit}, \binits{F.}},
\bauthor{\bsnm{Zeier}, \binits{P.}},
\bauthor{\bsnm{Kirsch}, \binits{P.}},
\bauthor{\bsnm{W{\"u}st}, \binits{S.}},
\bauthor{\bsnm{Wessa}, \binits{M.}}:
\batitle{Investigating individual stress reactivity: high hair cortisol predicts lower acute stress responses}.
\bjtitle{Psychoneuroendocrinology}
\bvolume{118},
\bfpage{104660}
(\byear{2020})
\end{barticle}
\endbibitem

%%% 62
\bibitem[\protect\citeauthoryear{Critchley}{2002}]{critchley2002electrodermal}
\begin{barticle}
\bauthor{\bsnm{Critchley}, \binits{H.D.}}:
\batitle{Electrodermal responses: what happens in the brain}.
\bjtitle{The Neuroscientist}
\bvolume{8}(\bissue{2}),
\bfpage{132}--\blpage{142}
(\byear{2002})
\end{barticle}
\endbibitem

%%% 63
\bibitem[\protect\citeauthoryear{Hou et~al.}{2015}]{hou2015eeg}
\begin{bchapter}
\bauthor{\bsnm{Hou}, \binits{X.}},
\bauthor{\bsnm{Liu}, \binits{Y.}},
\bauthor{\bsnm{Sourina}, \binits{O.}},
\bauthor{\bsnm{Tan}, \binits{Y.R.E.}},
\bauthor{\bsnm{Wang}, \binits{L.}},
\bauthor{\bsnm{Mueller-Wittig}, \binits{W.}}:
\bctitle{Eeg based stress monitoring}.
In: \bbtitle{2015 IEEE International Conference on Systems, Man, and Cybernetics},
pp. \bfpage{3110}--\blpage{3115}
(\byear{2015}).
\bcomment{IEEE}
\end{bchapter}
\endbibitem

%%% 64
\bibitem[\protect\citeauthoryear{Seo et~al.}{2010}]{seo2010stress}
\begin{botherref}
\oauthor{\bsnm{Seo}, \binits{S.-H.}},
\oauthor{\bsnm{Lee}, \binits{J.-T.}},
\oauthor{\bsnm{Crisan}, \binits{M.}}:
Stress and eeg.
Convergence and hybrid information technologies
\textbf{27}
(2010)
\end{botherref}
\endbibitem

%%% 65
\bibitem[\protect\citeauthoryear{Friedrich et~al.}{2023}]{friedrich2023your}
\begin{barticle}
\bauthor{\bsnm{Friedrich}, \binits{J.}},
\bauthor{\bsnm{Bareis}, \binits{A.}},
\bauthor{\bsnm{Bross}, \binits{M.}},
\bauthor{\bsnm{B{\"u}rger}, \binits{Z.}},
\bauthor{\bsnm{Cort{\'e}s~Rodr{\'\i}guez}, \binits{{\'A}.}},
\bauthor{\bsnm{Effenberger}, \binits{N.}},
\bauthor{\bsnm{Kleinhansl}, \binits{M.}},
\bauthor{\bsnm{Kremer}, \binits{F.}},
\bauthor{\bsnm{Schr{\"o}der}, \binits{C.}}:
\batitle{“how is your thesis going?”--ph. d. students’ perspectives on mental health and stress in academia}.
\bjtitle{Plos one}
\bvolume{18}(\bissue{7}),
\bfpage{0288103}
(\byear{2023})
\end{barticle}
\endbibitem

%%% 66
\bibitem[\protect\citeauthoryear{Gaillard}{1993}]{gaillard1993comparing}
\begin{barticle}
\bauthor{\bsnm{Gaillard}, \binits{A.W.}}:
\batitle{Comparing the concepts of mental load and stress}.
\bjtitle{Ergonomics}
\bvolume{36}(\bissue{9}),
\bfpage{991}--\blpage{1005}
(\byear{1993})
\end{barticle}
\endbibitem

%%% 67
\bibitem[\protect\citeauthoryear{Zhang et~al.}{2024}]{zhang2024good}
\begin{barticle}
\bauthor{\bsnm{Zhang}, \binits{X.}},
\bauthor{\bsnm{Zhao}, \binits{Z.}},
\bauthor{\bsnm{Sun}, \binits{J.}},
\bauthor{\bsnm{Ren}, \binits{J.}}:
\batitle{Good stress or bad stress? an empirical study on the impact of time pressure on doctoral students’ innovative behavior}.
\bjtitle{Frontiers in Psychology}
\bvolume{15},
\bfpage{1460037}
(\byear{2024})
\end{barticle}
\endbibitem

%%% 68
\bibitem[\protect\citeauthoryear{P{\'e}rez-Jorge et~al.}{2025}]{perez2025examining}
\begin{barticle}
\bauthor{\bsnm{P{\'e}rez-Jorge}, \binits{D.}},
\bauthor{\bsnm{Boutaba-Alehyan}, \binits{M.}},
\bauthor{\bsnm{Gonz{\'a}lez-Contreras}, \binits{A.I.}},
\bauthor{\bsnm{P{\'e}rez-P{\'e}rez}, \binits{I.}}:
\batitle{Examining the effects of academic stress on student well-being in higher education}.
\bjtitle{Humanities and Social Sciences Communications}
\bvolume{12}(\bissue{1}),
\bfpage{1}--\blpage{13}
(\byear{2025})
\end{barticle}
\endbibitem

%%% 69
\bibitem[\protect\citeauthoryear{Csikszentmihalyi et~al.}{2014}]{csikszentmihalyi2014flow}
\begin{botherref}
\oauthor{\bsnm{Csikszentmihalyi}, \binits{M.}},
\oauthor{\bsnm{Csikszentmihalyi}, \binits{M.}},
\oauthor{\bsnm{Abuhamdeh}, \binits{S.}},
\oauthor{\bsnm{Nakamura}, \binits{J.}}:
Flow.
Flow and the foundations of positive psychology: The collected works of Mihaly Csikszentmihalyi,
227--238
(2014)
\end{botherref}
\endbibitem

%%% 70
\bibitem[\protect\citeauthoryear{Peifer et~al.}{2014}]{peifer2014relation}
\begin{barticle}
\bauthor{\bsnm{Peifer}, \binits{C.}},
\bauthor{\bsnm{Schulz}, \binits{A.}},
\bauthor{\bsnm{Sch{\"a}chinger}, \binits{H.}},
\bauthor{\bsnm{Baumann}, \binits{N.}},
\bauthor{\bsnm{Antoni}, \binits{C.H.}}:
\batitle{The relation of flow-experience and physiological arousal under stress—can u shape it?}
\bjtitle{Journal of Experimental Social Psychology}
\bvolume{53},
\bfpage{62}--\blpage{69}
(\byear{2014})
\end{barticle}
\endbibitem

%%% 71
\bibitem[\protect\citeauthoryear{Ferreira}{2019}]{ferreira2019emotional}
\begin{barticle}
\bauthor{\bsnm{Ferreira}, \binits{S.O.}}:
\batitle{Emotional activation in human beings: Procedures for experimental stress induction}.
\bjtitle{Psicologia USP}
\bvolume{30},
\bfpage{180176}
(\byear{2019})
\end{barticle}
\endbibitem

%%% 72
\bibitem[\protect\citeauthoryear{Fahey et~al.}{2024}]{fahey2024importance}
\begin{barticle}
\bauthor{\bsnm{Fahey}, \binits{K.M.}},
\bauthor{\bsnm{Dermody}, \binits{S.S.}},
\bauthor{\bsnm{Cservenka}, \binits{A.}}:
\batitle{The importance of community engagement in experimental stress and substance use research with marginalized groups: lessons from research with sexual and gender minority populations}.
\bjtitle{Drug and alcohol dependence}
\bvolume{260},
\bfpage{111349}
(\byear{2024})
\end{barticle}
\endbibitem

%%% 73
\bibitem[\protect\citeauthoryear{{American Psychological Association}}{2017}]{apa2017ethics}
\begin{botherref}
\oauthor{\bsnm{{American Psychological Association}}}:
Ethical Principles of Psychologists and Code of Conduct.
Accessed: 2025-05-16
(2017).
\url{https://www.apa.org/ethics/code/index}
\end{botherref}
\endbibitem

%%% 74
\bibitem[\protect\citeauthoryear{Alonso-Fernandez et~al.}{2011}]{alonso2011quality}
\begin{barticle}
\bauthor{\bsnm{Alonso-Fernandez}, \binits{F.}},
\bauthor{\bsnm{Fierrez}, \binits{J.}},
\bauthor{\bsnm{Ortega-Garcia}, \binits{J.}}:
\batitle{Quality measures in biometric systems}.
\bjtitle{IEEE Security \& Privacy}
\bvolume{10}(\bissue{6}),
\bfpage{52}--\blpage{62}
(\byear{2011})
\end{barticle}
\endbibitem

\end{thebibliography}

\end{document}